\documentclass[11pt,a4paper,english]{article}

\usepackage{graphics}
\usepackage{color,graphicx}
\usepackage{amsmath,amsfonts,amssymb,amsbsy,times}
\usepackage{fontenc,url,path}
\usepackage{lscape,epsfig,epsf,psfrag}
\usepackage{stmaryrd}
\usepackage[usenames]{pstcol}
\usepackage{natbib}
\usepackage{stmaryrd}
\usepackage{xspace}
\usepackage{color,xcolor}
\usepackage{tikz}
\usepackage{graphicx}

\newcommand{\Esp}{\mathbb{E}}
\newcommand{\Var}{\mathbb{V}}
\newcommand{\Cov}{\mathbb{C}\text{ov}}

\newcommand{\Sigmabf}{\text{\mathversion{bold}{$\Sigma$}}}
\newcommand{\Psibf}{\text{\mathversion{bold}{$\Psi$}}}
\newcommand{\mubf}{\text{\mathversion{bold}{$\mu$}}}

\newcommand{\Sbf}{\mathbf{S}}
\newcommand{\Ybf}{\mathbf{Y}}
\newcommand{\Zbf}{\mathbf{Z}}
\newcommand{\Abf}{\mathbf{A}}
\newcommand{\Bbf}{\mathbf{B}}
\newcommand{\Bbfp}{\Bbf^{\prime}}
\newcommand{\Gbf}{\mathbf{G}}

\newcommand{\Wbf}{\mathbf{W}}

\newcommand{\Fbf}{\mathbf{F}}
\newcommand{\Rbf}{\mathbf{R}}
\newcommand{\Ubf}{\mathbf{U}}
\newcommand{\Ebf}{\mathbf{E}}
\newcommand{\Tbf}{\mathbf{T}}
\newcommand{\Ibb}{\mathbb{I}}
\newcommand{\Ibf}{\mathbf{I}}
\newcommand{\Jbf}{\mathbf{J}}
\newcommand{\RMSE}{\mbox{RMSE}}

\newcommand{\Lcal}{\mathcal{L}}
\newcommand{\FASeg}{{\tt FASeg}\xspace}

\def\1{1\!{\rm l}}

\newcommand{\SR}[2]{\textcolor{gray}{}\textcolor{black}{#2}}

\newcommand{\SRcom}[1]{\textcolor{cyan}{[{}]}}
\newcommand{\ELcom}[1]{\textcolor{green}{[{}]}}

\begin{document}

\title{A factor model approach for the joint segmentation with between-series correlation}

\author{Xavier Collilieux \thanks{IGN LAREG, Univ Paris Diderot, Paris, France} \and Emilie Lebarbier \thanks{AgroParisTech UMR518, Paris 5e and INRA UMR518, Paris 5e, FRANCE. e-mail:lebarbie@agroparistech.fr} \and St\'ephane Robin \thanks{AgroParisTech UMR518, Paris 5e and INRA UMR518, Paris 5e, FRANCE}}

\maketitle

Running title: Segmentation and factor model

\begin{abstract} We consider the segmentation of set of correlated time-series, the correlation being allowed to take an arbitrary form but being the same at each time-position. We show that encoding the dependency in a factor model enables us to use the dynamic programming algorithm for the inference of the breakpoints, which remains one the most efficient algorithm. We propose a model selection procedure to determine both the number of breakpoints and the number of factors. This proposed method is implemented in the \FASeg R package, which is available on the CRAN. We demonstrate the performances of our procedure through simulation experiments and an application to geodesic data is presented. 
\end{abstract}

{keyword:} Segmentation; Multivariate series; Factor model; Dynamic Programming; EM algorithm.

\section{Introduction}

\paragraph{{General segmentation problem}}

Segmentation methods aim at detecting abrupt changes --~called breakpoints~-- in
the distribution of a signal. Segmentation problems arise in many areas such as biology for the detection of chromosomal aberrations (\cite{PRL05,LJK05}), climatology for the detection of instrumental changes (\cite{CM04,Mal13}) or geodesy for the detection of changes in GPS location series either due to instrumental or to environmental changes as earth's crust shifts (\cite{William2003}). In many cases, multiple series (several patients (\cite{VB2010,NHPT2011}), meteorological stations or GPS receivers (\cite{gazeaux2015joint})) are observed simultaneously and dependence is likely to exist between them {due, e.g., to} probe effect in genomics  (\cite{PLB11}) or to spatial correlation in geodesy (\cite{D2006, Amiri2009}). \\

The literature on the segmentation of univariate series is too vast to cited here, the segmentation problem of multivariate series is more recent. We focus here on this latter problem, for which most works consider the detection of changes in the first-order or second-order structure of the series. Within this framework, we may still distinguish two types of segmentation. \\
The first one consists in the detection of breakpoints that are common to all the series and will be referred to as {\sl simultaneous segmentation} in the sequel. Changes in the mean of independent Gaussian series can be detected with a lasso-type approach \citep{VB2010} or generalized likelihood ratio tests \citep{ZS2008}, while penalized contrasts \citep{LT2006} or CUSUM-based binary segmentation algorithm \citep{AHHR2009,CF2012} have been proposed to detect changes in the covariance structure. Nonparametric approaches have also been proposed: \cite{MJ2014} optimize a weighted $L^2$ norm of the characteristic functions using sequential partitioning, while \cite{ACH2016} transform the problem into a least-square segmentation problem using an appropriate kernel. \cite{Cabri2016} provide a comparison study of several of these nonparametric breakpoint detection methods. \\
\SR{}{Not all simultaneous segmentation methods assume that a change actually occurs in each series at each breakpoint. \cite{ChF15} or \cite{WaS18} introduce some parsimony in the algorithm, so that only a fraction of the series displays a change at each detected breakpoint. In the Bayesian framework, appropriate priors can favor common breakpoints \cite[see][]{HCGA2016,DT2007}. } \\
As opposed to simultaneous segmentation, {\sl joint segmentation} assumes that the breakpoints are series-specific. \SR{}{In this case, segmenting all the series at the same time is of interest only when the series are affected by common effects, such as a probe effect in genomics (\cite{PLHRTR11}) or atmospheric effects in geodesy (\cite{gazeaux2015joint}).} \\ 
\SR{}{The choice between {\sl simultaneous} and {\sl joint} segmentation obviously depends on the application at hand. Simultaneous segmentation makes sense when (a fraction of) the series are expected to be affected by the same changes, whereas joint segmentation should be preferred when each series is expected to be affected by specific changes. Our contribution is about the later case.}

We consider here the joint segmentation in the mean of several series collected, say, in different locations and assuming that there is a correlation structure between them. This correlation is not assumed to be affected by the changes, so the problem is not to detect changes in the covariance structure as in aforementioned {\sl simultaneous} segmentation approaches. To our knowledge, \cite{PLB11} is the only work proposing a global estimation procedure in this setting. This reference highlights the fact that accounting for between-series correlation avoids false breakpoint detection. The global estimation procedure therein also avoids the opposite effect, that is to embed true breakpoints in the dependence structure when the correlation is estimated prior to segmentation.
The dependence structure we consider here differs substantially from \cite{PLB11} and the overlap between the two models corresponds to a very specific model. This point is detailed and discussed in Section \ref{sec:model}.   

\paragraph{Algorithmic issues in segmentation}
Segmentation methods have to deal with an inherent algorithmic complexity.
Indeed, the inference requires to search over the space of all possible segmentations, which is prohibitive in terms of computational time when performed in a naive way. The Dynamic Programming (DP) algorithm \citep{bellman_approximation_1961} and, its recent pruned versions \citep{killick_pelt,G15, Maidstone2016}, are the only algorithms that retrieve the exact solution in a fast way. Unfortunately, DP only applies when the contrast to be
optimized is additive with respect to the segments \citep[see][]{BP03,CM04,PRL05}. \SR{}{In the sequel, we refer to this condition as the additivity condition.} In the specific case of maximum likelihood inference, this condition is satisfied as soon as the data are independent from one segment to another. When dealing with the joint segmentation of multiple correlated series, the additivity condition is not satisfied due to the dependence between segments. The only case where the log-likelihood stays additive is the simultaneous segmentation framework. The goal of this paper is to propose an efficient maximum likelihood inference procedure for joint segmentation. More precisely, we use a generic representation of the between-series correlation that enables us to use DP.

\paragraph{Factor model, conditional independence and regularization}
In this article, we propose to encode the dependence between the series in a factor model. Factor models can describe any covariance structure between the series \SR{}{provided the number of factors is large enough}. The covariance results from the effect of a set of latent factors that affect all series at the same time. Because of the presence of latent factors, this representation suggests the use of an EM algorithm \citep{DLR77}, which takes a simple form in the Gaussian context. The interest of this representation is that the series are independent conditionally on the latent factors. As a consequence, at the M step of the EM algorithm, the contrast to be optimized turns out to be additive so DP applies for the segmentation parameter estimation. \\
Another property of the factor model is that it provides a parsimonious representation of the dependence structure when only few factors are used. The simplest modeling corresponds to one single factor and yields a uniform correlation. The same modeling principle has been successfully applied in another context where very little was known about the correlation structure \citep[see][]{FKC09}. Such a regularization is especially desirable when dealing with many series. Note that \cite{BCF17} also use a factor model to model the between-series dependence structure for breakpoint detection, but in the simultaneous segmentation context. 

\paragraph{Model selection} 
In segmentation, model selection traditionally refers to the choice of the number of breakpoints. A huge literature has been devoted to the subject: see e.g. \cite{Leb05,Lav05,ZhS07} for one series or \cite{CM04} for the multivariate case. In a similar context to ours, \cite{PLB11} adapted the mBIC criterion proposed by \cite{ZhS07} to the multivariate case and showed that this criterion achieves the best performances. On the other hand, BIC is the standard criterion to choose the number of factors in the factor model \citep{LoW04}.  We propose here a heuristic procedure combining these two BIC criteria in order to choose both the number of segments and the number of factors. 

\paragraph{Implementation} 
The proposed method has been implemented in the \FASeg R package, which is available on the CRAN.

\paragraph{Outline}
The article is organized as follows. In Section \ref{sec:model}, we present the proposed model which combines a segmentation model with a factor model. Section \ref{sec:graphical} gives a graphical model view-point \SR{}{that explains why the factor model structure of the dependence allows to get rid of it during the M-step of the EM algorithm, so that DP applies.} The EM algorithm providing the maximum likelihood estimates is described in Section~\ref{sec:procedure}. In Section~\ref{sec:modelselection}, we introduce a model selection procedure for both the number of segments and the number of factors. A simulation study is performed in Section~\ref{sec:simu} and in Section~\ref{sec:appli} we apply our method to geodesic data.

\section{Model} \label{sec:model}

We now define the joint segmentation (that is, breakpoint detection in the mean of multiple series) model that constitutes our general framework. In this model, the breakpoints are specific to each series and, at any given time, the series are correlated with the same correlation structure along time. This correlation is encoded in a factor model. In the next section, we will show how the representation in terms of factor model allows to breaks down some dependences.

\paragraph{Joint segmentation model}
We consider $M$ series with $n$ points each. We note $y_{tm}$ the observed signal of series $m$ at time $t$. The total number of observations is $N =n M$. The data are gathered in a matrix $\Ybf$ with dimension $[n \times M]$. For a given matrix $\Abf$, we denote by $\Abf_t$ its $t$-th row and by $\Abf^m$ its $m$-th column. Thus the column $\Ybf^m$ represents whole series $m$, while the row $\Ybf_t$ contains the observations at time $t$ in all series. We assume that the mean of the series $\Ybf^m = (Y_{tm})_{t=1..n}$ is subject to $K_{m}-1$ specific breakpoints at positions $(t_{k}^{m})_{k=0..K_m}$ (with convention $t^m_0=0$ and $t^m_{K_m}=n$) and is constant between two breakpoints or within the interval $I_{k}^{m} = \llbracket t_{k-1}^{m}+1, t_{k}^{m} \rrbracket$. We denote by $K=\sum_{m}^M K_{m}$ the total number of segments and $n_k^{m} = t_{k}^{m}- t_{k-1}^{m}$ the length of segment $k$ from series $m$ ($k=1,\ldots,K_m$). The segmentation model is written as follows: 
\begin{equation} \label{eq:model1}
Y_{tm} = \mu_{km} + F_{tm} \qquad \forall t \in I_k^{m},
\end{equation}
where the $M$-dimensional error vectors $\{\Fbf_t\}_{t=1..n}$ are supposed to be i.i.d. Gaussian, centered, with covariance matrix $\Sigmabf$. Observe that simultaneous segmentation corresponds to the special case where $t^m_k \equiv t_k$ and $I^m_k \equiv I_k$.

\paragraph{Factor model} 
We first remind that any symmetric definite positive $M$-dimensional matrix can be rewritten as 
\begin{eqnarray} \label{eq:variance_decomposition}
\Sigmabf = \Bbf \Bbfp + \Psibf
\end{eqnarray}
\SR{}{where $\Bbf = (b_{mq})$ is a fixed $[M \times Q]$ matrix with $Q =M-1$ and $\Psibf$ is diagonal.} 

\SR{}{The factor model imposes a specific structure to the covariance matrix, assuming that $Q \leq M-1$. Writing $\Sigmabf$ as in \eqref{eq:variance_decomposition} also provides a latent factor interpretation of the covariance structure.} Formally, let us consider i.i.d. random centered Gaussian vectors $\Zbf_t = (Z_{tq})$ with covariance matrix $\Ibf_Q$ and i.i.d. random centered Gaussian vectors $\Ebf_t = (E_{tm})$ with diagonal covariance matrix $\Psibf$ and independent from the $\Zbf_t$. The error term from Model \eqref{eq:model1} has the same distribution as a linear combination of the $Z_{tq}$ plus $E_{tm}$:
\begin{eqnarray*}
\forall t, m, \quad F_{tm} \overset{\Delta}{=} \sum_{q=1}^Q Z_{tq} b_{mq} + E_{tm} 
\qquad \Leftrightarrow \qquad 
\forall t, \Fbf_t \overset{\Delta}{=} \Zbf_t \Bbfp + E_t
\end{eqnarray*}
In this representation, the vectors $\Zbf_t$ are the latent vectors that capture all the dependence between the series. At each time $t$, the series $\Ybf_t$ are independent conditionally on $Z_t$. Note that the assumption $\Var(\Zbf_t) = \Ibf_Q$ is necessary for identifiability reasons. \SR{}{This model is similar to the factor model proposed in \cite{BaN02}.}

\paragraph{Latent factor version of Model \eqref{eq:model1}} 
With the previous representation, Model (\ref{eq:model1}) can be rewritten as a mixed linear model:
\begin{equation} \label{eq:model}
Y_{tm} = \mu_{km} + \sum_{q=1}^{Q} Z_{tq} b_{mq}+E_{tm} \qquad \forall
t \in I_k^{m}.
\end{equation}
We denote by $\1_l$ the $[l \times 1]$ vector with all coordinates equal to one. For a collection of column vectors $(C_i)_{i=1,\ldots,I}$, we denote by $\text{Bloc}\left[ C_i \right]_i$ the block diagonal matrix with blocks $(C_i)_{i=1,\ldots,I}$. 
Hence, the previous linear model can be rephrased in a matrix form: 
\begin{equation} \label{eq:modelMat}
\Ybf=\Tbf \mubf +\Zbf \Bbfp+\Ebf,
\end{equation}
where, $\Ybf$ stands for the $[n \times M]$ observed data matrix, $\Tbf$ is the $[n\times K]$ incidence matrix of breakpoints: ${T^m} = \text{Bloc} \left[ \1_{n_{k}^{m}} \right]_{k=1,\ldots,K_m}$ and ${\Tbf} = \left[ T^1 \ T^2 \ \ldots \ T^M \right]$, $\mubf$ is the $[K\times M]$ mean matrix (and $\mu_{k}^m$ the mean of segment $k$ in series $m$) such that ${\mubf^{m}}= \left[\mu^m_1  \ \mu^m_2  \ \ldots  \ \mu^m_{K_m} \right]'$ and $ {\mubf}  = \text{Bloc} \left[\mu^m \right]_{m=1,\ldots,M}$, $\Zbf$ has size $[n\times Q]$ and $\Bbf$ has size $[M \times Q]$, and $\Ebf$ has size $[n\times M]$, \SR{}{each of its row being a centered Gaussian with a diagonal covariance matrix $\Psibf$}. The unknown parameters of this model are gathered into
$$
\phi=(\Tbf,\mubf,\Psibf,\Bbf).
$$
The main difference between Model \eqref{eq:modelMat} and a classical mixed linear model is that both the incidence matrix $\Tbf$ and the factor matrix $\Bbf$ are unknown. 

\paragraph{Comparison with \cite{PLB11}} 
In this paragraph, we discuss the differences between the model proposed in \cite{PLB11} and the model we consider, which mostly lie in the dependence structure. To emphasize the difference, we stack the $M$ vectors $({\bf Y}^m)_m$ into a one single vector defined as
\begin{equation*}
\text{vec}(\bf{Y}) = \left[\begin{array}{c} 
{\bf Y}^1 \\ {\bf Y}^2 \\ \vdots \\ {\bf Y}^M
\end{array} \right].
\end{equation*}
The model proposed by \cite{PLB11} writes
\begin{equation*}
\text{vec}(\bf{Y}) = \Tbf \mubf + \Zbf \Ubf + \Ebf,
\end{equation*}
where $\Ebf$ is a centered Gaussian $[N \times 1]$ vector with diagonal covariance matrix $\Rbf$, $\Zbf$ is the $[N \times n]$ incidence matrix which associates each entry of $\Ybf$ with its respective observation time $t \in [1; n]$, $\Ubf$ is a centered Gaussian $[n \times 1]$ vector with covariance matrix $\Gbf$ and $\Ubf$ and $\Ebf$ are supposed independent. Because of the forms of $\Zbf$ and $\Ubf$, each random effect $U_t$ affects all series at time $t$ in the same way. The two models yield the following covariance structures:
$$
\begin{array}{l|c|c}
 & \quad \text{\cite{PLB11}} \quad ~ & \qquad \text{Our model} \qquad ~ \\
 \hline 
 \Var(\text{vec}(\bf{Y})) & \Ibf_M \otimes \Gbf + \Rbf & \SR{}{(\Bbf \Bbfp) \otimes  \Ibf_n + \Psibf \otimes  \Ibf_n =: \tilde{\Bbf}  \otimes  \Ibf_n + \tilde{\Rbf}} \\
 \Var(Y_{tm}) & G_{tt} + R_{tm,tm} &  \SR{}{\tilde{B}_{m m}+\tilde{R}_{tm,tm}} \\
 \Cov(Y_{tm}, Y_{tm'}) & G_{tt} & \SR{}{\tilde{B}_{m m'}} \\
 \Cov(Y_{tm}, Y_{t'm}) & G_{tt'} & 0
\end{array}
$$ 
As a consequence, in \cite{PLB11}, at a given time $t$, the correlation between all the series is the same, as $G_{tt'}$ does not depend on $(m,  m'$). Conversely, in our model this covariance \SR{}{$\tilde{B}_{m m'}$} depends on each pair of series, without any specific structure: this why we name it 'arbitrary'. Moreover, \cite{PLB11} can account for correlation between different times (through the term $G_{tt'}$), whereas our model does not. \\ 
As a consequence, the two modelings do not overlap in general. The only case that we found where the models coincide is quite specific:
\begin{description}
 \item[\cite{PLB11}:] $\Gbf = \sigma^2_U \Ibf_n$ (no time dependence),  $\Rbf = \Psibf \otimes \Ibf_n$ (constant residual variance within each series);
 \item[Our model:] $\Bbf = \sigma_U \mathbf{1}_M$ (one single factor with constant coefficient) so $\Sigmabf = \sigma^2_U \Jbf_M + \Psibf$.
\end{description}
In both case, we get $\Var(Y_{tm}) = \sigma^2_U + \psi_{mm}$, $\Cov(Y_{tm}, Y_{tm'}) = \sigma^2_U$ and $\Cov(Y_{tm}, Y_{t'm}) = 0$.

\section{A graphical model point-of-view} \label{sec:graphical}
In the present section, we explain how the use of the latent factors enables us to use \SR{}{DP to retrieve the maximum-likelihood segmentation in an efficient way.} 
We illustrate this point based on both the likelihood and the graphical representation (given in Figure \ref{Fig:GraphModel}) of Models \eqref{eq:model1} and \eqref{eq:model}.

\newcommand{\ns}{3.5em}
\newcommand{\veu}{1.5*\ns} 
\newcommand{\heu}{2.2*\ns} 
\newcommand{\vseg}{0.5*\ns} 
\newcommand{\hseg}{0.25*\heu} 
\newcommand{\seg}[3]{\draw[dashed] (#1*\heu-\hseg, #3*\veu-\vseg) -- (#1*\heu-\hseg, #3*\veu+\vseg) -- (#2*\heu+\hseg, #3*\veu+\vseg) -- (#2*\heu+\hseg, #3*\veu-\vseg) -- (#1*\heu-\hseg, #3*\veu-\vseg);}
\newcommand{\segleft}[3]{\draw[solid] (#1*\heu-\hseg, #3*\veu+\vseg) -- (#2*\heu+\hseg, #3*\veu+\vseg) -- (#2*\heu+\hseg, #3*\veu-\vseg) -- (#1*\heu-\hseg, #3*\veu-\vseg);}
\newcommand{\segright}[3]{\draw[solid] (#2*\heu+\hseg, #3*\veu-\vseg) -- (#1*\heu-\hseg, #3*\veu-\vseg) -- (#1*\heu-\hseg, #3*\veu+\vseg) -- (#2*\heu+\hseg, #3*\veu+\vseg);}
\tikzstyle{hidden}=[minimum width=\ns, inner sep=0]
\tikzstyle{observed}=[minimum width=\ns, inner sep=0]
\tikzstyle{eliminated}=[minimum width=\ns, color=gray!125, inner sep=0]
\tikzstyle{empty}=[]
\tikzstyle{arrow}=[->, >=latex, line width=1pt, ]
\tikzstyle{edge}=[-, line width=1pt]
\tikzstyle{lightarrow}=[->, >=latex, line width=1pt, fill=gray!80, color=gray!80]

\begin{figure}
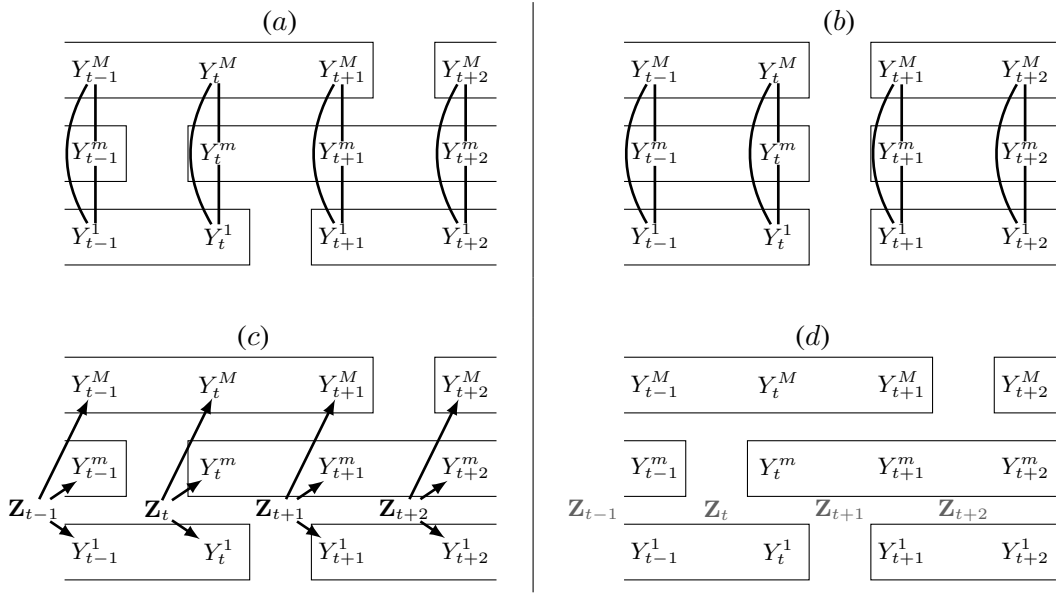

  \begin{center}
  \hspace{-0.075\textwidth}
  \begin{tabular}{r|r}
   \begin{tabular}{c}
   ($a$) \\
    {\tiny   \begin{tikzpicture}
  \input{Fig_Model1-nodes.tex}
  \segleft{0}{2}{2}; \segright{3}{3}{2}
  \segleft{0}{0}{1}; \segright{1}{3}{1}
  \segleft{0}{1}{0}; \segright{2}{3}{0}
  \draw[edge] (Y1tm1) to (Ymtm1);   
  \draw[edge] (Y1tm1) to [bend left](YMtm1);  
  \draw[edge] (Ymtm1) to (YMtm1);  
  \draw[edge] (Y1t) to (Ymt);   
  \draw[edge] (Y1t) to [bend left](YMt);  
  \draw[edge] (Ymt) to (YMt);  
  \draw[edge] (Y1tp1) to (Ymtp1);   
  \draw[edge] (Y1tp1) to [bend left](YMtp1);  
  \draw[edge] (Ymtp1) to (YMtp1);  
  \draw[edge] (Y1tp2) to (Ymtp2);   
  \draw[edge] (Y1tp2) to [bend left](YMtp2);  
  \draw[edge] (Ymtp2) to (YMtp2);  
  \end{tikzpicture}} 
   \end{tabular}
   &
   \begin{tabular}{c}
   ($b$) \\
    {\tiny   \begin{tikzpicture}
  \input{Fig_Model1-nodes.tex}
  \segleft{0}{1}{2}; \segright{2}{3}{2}
  \segleft{0}{1}{1}; \segright{2}{3}{1}
  \segleft{0}{1}{0}; \segright{2}{3}{0}
  \draw[edge] (Y1tm1) to (Ymtm1);   
  \draw[edge] (Y1tm1) to [bend left](YMtm1);  
  \draw[edge] (Ymtm1) to (YMtm1);  
  \draw[edge] (Y1t) to (Ymt);   
  \draw[edge] (Y1t) to [bend left](YMt);  
  \draw[edge] (Ymt) to (YMt);  
  \draw[edge] (Y1tp1) to (Ymtp1);   
  \draw[edge] (Y1tp1) to [bend left](YMtp1);  
  \draw[edge] (Ymtp1) to (YMtp1);  
  \draw[edge] (Y1tp2) to (Ymtp2);   
  \draw[edge] (Y1tp2) to [bend left](YMtp2);  
  \draw[edge] (Ymtp2) to (YMtp2);  
  \end{tikzpicture}} 
   \end{tabular}
   \\ ~\\ 
   \begin{tabular}{c}
   ($c$) \\
    {\tiny   \begin{tikzpicture}
  \input{Fig_Model1-nodes.tex}
  \segleft{0}{2}{2}; \segright{3}{3}{2}
  \segleft{0}{0}{1}; \segright{1}{3}{1}
  \segleft{0}{1}{0}; \segright{2}{3}{0}
  \node[hidden] (Ztm1) at (-.5*\heu, .5*\veu) {\footnotesize $\Zbf_{t-1}$};
  \node[hidden] (Zt) at (.5*\heu, .5*\veu) {\footnotesize $\Zbf_{t}$};
  \node[hidden] (Ztp1) at (1.5*\heu, .5*\veu) {\footnotesize $\Zbf_{t+1}$};
  \node[hidden] (Ztp2) at (2.5*\heu, .5*\veu) {\footnotesize $\Zbf_{t+2}$};
  \draw[arrow] (Ztm1) to (Y1tm1);   
  \draw[arrow] (Ztm1) to (Ymtm1);   
  \draw[arrow] (Ztm1) to (YMtm1);   
  \draw[arrow] (Zt) to (Y1t);   
  \draw[arrow] (Zt) to (Ymt);   
  \draw[arrow] (Zt) to (YMt);   
  \draw[arrow] (Ztp1) to (Y1tp1);   
  \draw[arrow] (Ztp1) to (Ymtp1);   
  \draw[arrow] (Ztp1) to (YMtp1);   
  \draw[arrow] (Ztp2) to (Y1tp2);   
  \draw[arrow] (Ztp2) to (Ymtp2);   
  \draw[arrow] (Ztp2) to (YMtp2);   
  \end{tikzpicture}} 
   \end{tabular}
   &
   \begin{tabular}{c}
   ($d$) \\
    {\tiny   \begin{tikzpicture}
  \input{Fig_Model1-nodes.tex}
  \node[eliminated] (Ztm1) at (-.5*\heu, .5*\veu) {\footnotesize $\Zbf_{t-1}$};
  \node[eliminated] (Zt) at (.5*\heu, .5*\veu) {\footnotesize $\Zbf_{t}$};
  \node[eliminated] (Ztp1) at (1.5*\heu, .5*\veu) {\footnotesize $\Zbf_{t+1}$};
  \node[eliminated] (Ztp2) at (2.5*\heu, .5*\veu) {\footnotesize $\Zbf_{t+2}$};
  \segleft{0}{2}{2}; \segright{3}{3}{2}
  \segleft{0}{0}{1}; \segright{1}{3}{1}
  \segleft{0}{1}{0}; \segright{2}{3}{0}
  \end{tikzpicture}} 
   \end{tabular}
  \end{tabular}
  \end{center}
  \caption{Graphical representation of the considered models. \SR{}{Edges and arrows depict dependences as defined in \cite{Lau96}}. The dashed blocks represent the segments. For the sake of readability, $Y_{tm}$ is denoted $Y_t^m$. ($a$) Model \eqref{eq:model1}; ($b$) special case of Model \eqref{eq:model1} when the breakpoints are the same in all series (simultaneous segmentation); ($c$) Model \eqref{eq:model}; ($d$)  conditional dependencies of $\Ybf|\Zbf$ in Model \eqref{eq:model}. \label{Fig:GraphModel}}
\end{figure}

\paragraph{Model \eqref{eq:model1}}
The likelihood of Model \eqref{eq:model1} is given by
$$
-2 \log \Lcal(\Ybf; \phi) = N \log{(2 \pi)} + n \log{(|\Sigmabf|)} +\sum_{t=1}^n \|\Ybf_t - \mubf_t\|^2_{\Sigmabf^{-1}}
$$
where $\mubf_t$ is the vector of the means of all series at time $t$. Because $\Sigmabf$ is not diagonal the observations from all series at a given time $t$ are not independent. As consequence time-overlapping segments from different series are not independent either. This is illustrated in Figure \ref{Fig:GraphModel} ($a$) where the left segment of series $M$ is correlated to the right segment of series 1 and $m$. This dependence hampers the use of DP. Note that, even in the case where $\Sigmabf$ is known, the problem would remain. \\
DP only applies in the case where the breakpoints are the same in all series (Figure \ref{Fig:GraphModel} ($b$)), which corresponds to simultaneous segmentation.

\paragraph{Model \eqref{eq:model}}
The graphical representation of the factor model \eqref{eq:model} is given in Figure \ref{Fig:GraphModel} ($c$), which reminds that the dependence between all series at time $t$ is induced only by $\Zbf_t$. The likelihood of this model is 
$
\log \Lcal(\Ybf,\Zbf; \phi) = \sum_t \log \Lcal(\Zbf_t) + \log \Lcal(\Ybf_t|\Zbf_t; \phi)
$
where only the second term depends on the parameter $\phi$. We further have
\begin{eqnarray} \label{eq:lik_cond}
-2  \sum_t \log \Lcal(\Ybf_t|\Zbf_t; \phi) 
=  N  \log{(2 \pi)}+ n
\log{(|\Psibf|)} +\sum_{t=1}^n \|\Ybf_t - \mubf_t - \Zbf_t
\Bbfp\|^2_{\Psibf^{-1}} 
\end{eqnarray}
Because $\Psibf$ is diagonal, the last term is written as
\begin{eqnarray*}
\sum_{t=1}^n \|\Ybf_t - \mubf_t - \Zbf_t \Bbfp\|^2_{\Psibf^{-1}}  
& = &
\sum_{m=1}^M \sum_{k=1}^{K_m} \sum_{t=t_{k-1}^m+1}^{t_k^m} (Y_{tm}- \mu_{km} - \sum_q Z_{tq} b_{qm})^2 / \psi_{mm}
\end{eqnarray*}
where the third sum involves all the observations from the $k$-th segment from series $m$, and only them. $\sum_{t=1}^n \|\Ybf_t - \mubf_t - \Zbf_t \Bbfp\|^2_{\Psibf^{-1}}$ is therefore additive with respect to the segments. If $\Zbf$ was observed (see Figure \ref{Fig:GraphModel} ($d$)), DP could estimate the segmentation parameters $(\Tbf, \mubf)$. As $\Zbf$ is not observed we will use an EM algorithm that will take advantage of this property.

\section{Estimation using the EM algorithm} \label{sec:procedure}
As mentioned above, we resort to maximum likelihood procedure to infer the parameter $\phi$, the number of factors $Q$ and the number of segments $K$ being fixed. We remind that the dependence structure is encoded in the latent factors $\Zbf$. The EM algorithm \citep{DLR77} is a classical tool for maximum likelihood inference in presence of latent (or missing) variables (see, e.g. \cite{VD00} for linear mixed models or \citep{RuT82} for factor models). We remind that the key quantity in the EM algorithm is the expectation of the log-likelihood of the complete data conditionally on the observed data $Y$, 
namely $Q(\phi; \phi^{(h)}):=\Esp_{\phi^{(h)}}\left\{\log \Lcal(\Ybf,\Zbf; \phi) | \Ybf \right\}$. 
As seen in the previous section, only the conditional distribution of $\Ybf$ given $\Zbf$ needs be considered to estimate the $\phi$. According to equation \eqref{eq:lik_cond}, $-2 Q(\phi; \phi^{(h)})$ is
$$
n \log |\Psibf|+\sum_{t=1}^n \left [ \|\Ybf_t - \mubf_{t} - \widehat{\Zbf}_t^{(h)}
\Bbfp\|^2_{\Psibf^{-1}} +\text{Tr} \left( \Bbfp
  \Psibf^{-1} \Bbf \Wbf_t^{(h)} \right) \right ],
$$
where $\phi^{(h)}$ is the value of $\phi$ at iteration $(h)$, $\Esp_{\phi}\{\cdot\}$ is the expectation calculated
with $\phi$ as the parameter value and $\Var_{\phi}\{\cdot\}$
the corresponding variance, $\widehat{\Zbf}_t^{(h)} = \Esp_{\phi^{(h)}}\left\{\Zbf_t |\Ybf\right\}$,
and $\Wbf_t^{(h)}=\Var_{\phi^{(h)}}\left\{\Zbf_t | \Ybf\right\}$.
$\text{Tr}(A)$ stands for the trace of matrix $A$ and $|A|$ for its determinant. 

The EM algorithm is iterative and each iteration consists of two steps: the E-step and the M-step. At iteration $(h+1)$, we have
\begin{description}
 \item[E-step:] it consists in the calculation of the conditional expectation $Q(\phi; \phi^{(h)})$ which requires the conditional moments $\widehat{\Zbf}$ and $\Wbf$. Denoting $\widetilde{\Ybf}_t^{(h)}=\Ybf_t-\mubf_{t}^{(h)}$, we get
 \begin{eqnarray*}
 \widehat{\Zbf}_t^{(h+1)} = \widetilde{\Ybf}_t^{(h)}  {\Psibf^{(h)}}^{-1} \Bbf^{(h)} \Wbf_t^{(h)}, 
 \qquad 
 \Wbf_t^{(h+1)} = \left (\Ibf_Q+ \Bbf'^{(h)} {\Psibf^{(h)}}^{-1} \Bbf^{(h)}  \right )^{-1}.
 \end{eqnarray*}
 \item[M-step:] it consists in the estimation of the parameters by maximizing the obtained conditional expectation. We get
 \begin{eqnarray*}
 \Bbf^{(h+1)} & = & 
 \left[\sum_{t=1}^n (\Ybf_t-\mubf_{t}^{(h)})' \widehat{\Zbf}_t^{(h+1)} \right]
 \left [ \sum_{t=1}^n (\widehat{\Zbf}_t'^{(h+1)} \widehat{\Zbf}_t^{(h+1)}+ \Wbf_t^{(h+1)})\right]^{-1}, \\
 \Psibf^{(h+1)}&=&\arg \max_{\Psibf} Q(\phi; \Tbf^{(h)}, \mubf^{(h)}, \Psibf, \Bbf^{(h+1)})= \frac{1}{n} \sum_{t=1}^n
 (\Ybf_t-\mubf_{t}^{(h)})' \Ebf_t^{(h)},
 \end{eqnarray*}
 where $\Ebf_t^{(h)}=\Ybf_t-\mubf_{t}^{(h)}-\widehat{\Zbf}_t^{(h+1)}\Bbf'^{(h+1)}$.
 In the case where $\Psibf=\sigma^2 \Ibf_M$, as we consider in the simulation study and the
 application, we get
 $
 \sigma^{2,(h+1)} =\frac{1}{N} \sum_{t=1}^n \left [ \Ebf_t^{(h)} \Ebf_t'^{(h)}+ \text{Tr} \left( \Bbf'^{(h)} \Bbf^{(h)}  \Wbf_t^{(h+1)} \right)\right ]
 $
 and $\Psibf^{(h+1)}=\sigma^{2,(h+1)} \Ibf_M$. \\
 As for the segmentation parameters $\Tbf \mubf$, $\Psibf$ being diagonal, we have that
 \begin{eqnarray} \label{eq:RemoveDep}
 \left\{ \Tbf^{(h+1)},\mubf^{(h+1)} \right\}  
 & = & \arg \min_{\Tbf,\mubf} \sum_{m=1}^M \sum_{k=1}^{K_m} 
 \sum_{t=t_{k-1}^m+1}^{t_k^m} (\breve{Y}_{tm}- \mu_{km})^2 / \psi_{mm}^{(h+1)}. 
 \end{eqnarray}
 where $\breve{\Ybf}_t=\Ybf_t- \widehat{\Zbf}_t^{(h+1)} \Bbf'^{(h+1)}$. This turns into a classical segmentation problem for which DP applies, the cost of segment $\llbracket t_{k-1}^m+1 , t_k^m \rrbracket$ from series $m$ being $\sum_{t=t_{k-1}^m+1}^{t_k^m} (\breve{Y}_{tm}- \mu_{km})^2 / \psi_{mm}^{(h+1)}$. In the case where $\Psibf=\sigma^2 \Ibf_M$ (homoscedastic noise case), the quantity to be minimized is a residual sum of squares. In the heteroscedastic noise case, the sum of squares to be minimized is a weighted version of it. In practice, we use the two-stage DP proposed by \cite{PLB11,PLHRTR11}, which is a fast version of DP dedicated to the joint segmentation of multiple series. 
\end{description}

\paragraph{Time complexity} As mentioned in \cite{PLB11}, the two-stage DP algorithm reduces the algorithmic complexity from $\mathcal{O}(n^2 M^2 K)$, which one would get with the classical DP, to $\mathcal{O}(Kn^2+K^2M)$. The complexity of all other steps is at most $\mathcal{O}(nM)$ so the global complexity of the algorithm is this of the segmentation part.

\section{Model selection} \label{sec:modelselection}
Both the number of factors $Q$ and the number of segments $K$ have to be estimated. This joint model selection issue is not standard and the difficulty arises from the different nature of the parameters at hand. Indeed the likelihood function is continuous with respect to the loadings $\Bbf$ of the latent vectors, so the classical framework for the BIC approximation holds. This is not true for the segmentation parameters and a specific BIC approximation needs to be derived, as observed and proposed by \cite{ZhS07}. Furthermore, the two resulting criteria do not share the same form, so they can not be combined into a single criterion. We propose here a two-step heuristic to select these two parameters.

As recalled in \cite{LoW04}, BIC is the most popular criterion to choose the number of factors in a factor model. Following this line, we propose to use this criterion to choose the number of factors $Q$ for a fixed number of segments $K$. We define:
$$
BIC_K(Q) = 2 \log \Lcal(\widehat{\Tbf \mubf}_K, \widehat{\Sigmabf}_Q)-D_Q \log{(n)},
$$
where ${\Sigmabf}_Q$ is the covariance matrix for a given $Q$, ${\Tbf \mubf}_K$ are the segmentation parameters for a given $K$, $\log \Lcal(\widehat{\Tbf \mubf}_K, \widehat{\Sigmabf}_Q)$ is the maximized log-likelihood for $K$ segments and $Q$ factors, and $D_Q$ is the number of parameters in a model with $Q$ factors, that is: $D_Q=Q(2 M -Q+1)/2+1$ in the case where $\Psibf=\sigma^2 \Ibf_M$. Indeed $\Bbf$ contains $MQ$ parameters but, due to orthogonality constraints, only $M Q - Q(Q-1)/2$ of them need to be estimated. The additional parameter is $\sigma^2$.

Symmetrically, for a given $Q$, we select the number of segments $K$ using the modified BIC criterion proposed by \cite{ZhS07} and adapted to the joint segmentation of multiple series by \cite{PLB11}. This criterion can be written as follows:
\begin{eqnarray*}
mBIC_Q(K) 
& = & 
\left ( \frac{K-M}{2} \right )\log{\left(\frac{SS_{\text{all}}}{2}\right )} 
+\left ( \frac{N-K}{2}+1 \right) \log{\left(1+\frac{SS_{\text{bg}}(\widehat{t})}{SS_{\text{wg}}(\widehat{t})}\right )}\\
&& + \log{\left [\Gamma \left (\frac{N-K}{2}+1 \right)\right ]}  -\frac{1}{2} \sum_{m=1}^M \sum_{k=1}^{K_m} \log
{\widehat{n}^m_k}-\left (K-M \right ) \log{(N)},
\end{eqnarray*}
with $SS_{\text{wg}}(\widehat{t}) = \sum_{t=1}^n (\Ybf_t-\widehat{\mubf}_{t}) \widehat{\Sigmabf}_Q^{-1}(\Ybf_t-\widehat{\mubf}_{t})'$, $SS_{\text{all}} = \sum_{t=1}^n (\Ybf_t-\bar{\Ybf}) \widehat{\Sigmabf}_Q^{-1}(\Ybf_t-\bar{\Ybf})'$, $SS_{\text{bg}}(\widehat{t}) = SS_{\text{all}}-SS_{\text{wg}}(\widehat{t})$ and where $\widehat{n}^m_k$ is the length of segment $k$ in series $m$ ($\widehat{n}^m_k=\widehat{t}^m_k-\widehat{t}^m_{k-1}$), $\widehat{\mubf}_{t}=\widehat{\mubf}_{k(t)}$ is a vector of size $M$ with the component $m$ is $\bar{y}_{mk}=(\widehat{n}^m_k)^{-1} \sum_{t=\widehat{t}^m_{k-1}+1}^{\widehat{t}^m_{k}} y_m(t)$ if $t \in \widehat{I}^m_k$ and $\bar{\Ybf}= (\sum_{t,m} Y_{tm})/N \ \mathbf{1}_M$. 
Part of the simulation study of \cite{PLB11} is devoted to the comparison of different model selection criteria \citep[cf][]{Leb05,Lav05,ZhS07,CM04} adapted to the joint segmentation context. The main conclusion is that the modified BIC criterion performs best. 

We end-up the following two-stage heuristic: choose the best $Q$ for
each $K$, then select the best $K$ among them:
\begin{eqnarray*}\widehat{Q}_K = \arg\max_Q BIC_K(Q), 
\qquad
\widehat{K}_{\widehat{Q}_K}= \arg\max_K mBIC_{\widehat{Q}_K}(K),
\qquad 
\widehat{Q} = \widehat{Q}_{\widehat{K}}.
\end{eqnarray*}

\section{Simulation study} \label{sec:simu}
In this section, we illustrate the importance of accounting for the dependence and we study the behavior of our model selection heuristic for $K$ and $Q$ and its impact on the estimation of all parameters.

\subsection{Simulation design and quality criteria}

\paragraph{Simulation design}

We consider different number of series $M \in \{5,10\}$ with different lengths $n \in \{50,100\}$. For each series $\Ybf^m$, the number of breakpoints ($K_m-1$) is Poisson distributed with mean $\bar{k}$ ($\bar{k}=3$ for $n=50$ and $\bar{k}=5$ for $n=100$) and their positions are uniformly distributed. Because the breakpoint locations are independent from one series to another, the comparison with simultaneous breakpoint detection methods is irrelevant. The mean value within each segment alternates between $0$ and one value picked in $\{-2, -1,+1,
+2\}$. We consider different residual standard deviations $\sigma \in \{.2,.5,1\}$ and a spatial-type correlation between series: distances $d$ are simulated as distances between Gaussian bivariate random vectors and the covariance matrix $\Sigmabf$ is defined as $\Sigma_{mm'}= \sigma^2 ((1-\alpha) \rho^{d_{mm'}} +\alpha \Ibb\{m = m'\})$ with $\alpha=0.2$ and $\rho \in \{0.2,0.8\}$. As a consequence, the true value of $Q$ is $M-1$. The parameter $\rho$ controls the intensity of the dependence between series. Each configuration is simulated 100 times.\\
We performed additional separate simulations to study the robustness of the proposed method to ($i$) non-normality and ($ii$) variance heterogeneity. For these studies, we consider the design with $M=10$, $n=100$, $\sigma=0.2$ and $\rho=0.8$. For ($i$), we consider a heavy tailed distribution for the errors $\{\Fbf_t\}_t$, namely a multivariate student distribution with covariance matrix $\Sigmabf$ and degrees of freedom $\nu=\{50,10,6,3\}$ ($\nu=50$ being the closest Gaussian case). For ($ii$), we simulate heterogeneous covariance matrices $\{\Sbf_t\}_t$ as i.i.d. copies from a $\mathcal{W}(\Sigmabf, \nu)$ Wishart distribution, taking $\nu = \{10, 100, 1000\}$. We then simulate independent, centered, Gaussian residual vectors $\Fbf_t$ with respective variance $\Sbf_t$.

\paragraph{Quality criteria} To assess the quality of the estimation of the covariance matrix, we use the root mean squared distance between the true parameter and its estimate: $\RMSE(\Sigmabf) =   \left[M^{-2} \sum_{m, m'=1}^{M}  (\widehat{\Sigma}_{mm'} -  \Sigma_{mm'})^2\right]^{1/2}$. For the segmentation parameters, 
we consider both the proportion of erroneously detected breakpoints among detected breakpoints (false positive rate, FPR) and the proportion of detected true breakpoints among true breakpoints (true positive rate, TPR). A perfect segmentation results in null FPR and TPR equals to 1. For each configuration we consider the average of these criteria over the 100 simulations. \\
\SR{}{We also compare our results to the results obtained when the dependence is not taken into account in the segmentation procedure, which corresponds to $Q=0$.}

\subsection{Results}

Only the results with $M=10$ and $n=100$ are presented as the results for the other configurations lead to same conclusions. In the graphs displayed in this section, we label  the true parameters with '$^*$' and the estimated ones with '$est$'.

\paragraph{Accounting for the dependence}

\begin{figure}
\begin{center}
\includegraphics[scale=0.6]{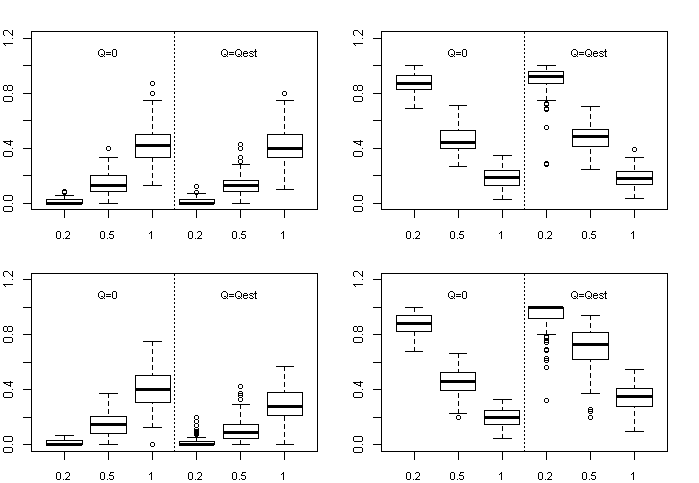}
\caption{FPR on the left and TPR on the right for $\rho=0.2$
(top) and $\rho=0.8$ (bottom) using $\widehat{K}$. We distinguish
the cases $Q=\widehat{Q}$ (denoted Qest on the graph) and $Q=0$
(e.g. the segmentation only). $x$-axis: $\sigma$. {Box: first, second and third quartiles.}}
\label{fig:dependency}
\end{center}
\end{figure}

Figure \ref{fig:dependency} compares the selected segmentation when the dependence is considered ($Q=\widehat{Q}$) or not ($Q=0$). Whatever the difficulty of the detection problem (different values of $\sigma$), accounting for the dependence increases the
performance of the segmentation (smaller FPR and larger TPR). The performances most differ when the dependence is high ($\rho=0.8$, bottom of the figure). \SR{}{Differences in terms of FDR and TPR are significant except for $\sigma=0.2$.} 

\begin{figure}
\begin{center}
\includegraphics[width=0.45\textwidth]{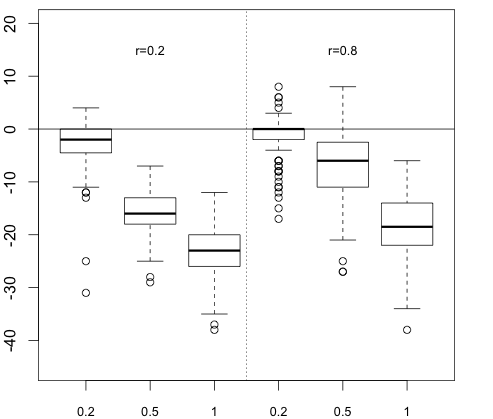}
\includegraphics[width=0.45\textwidth]{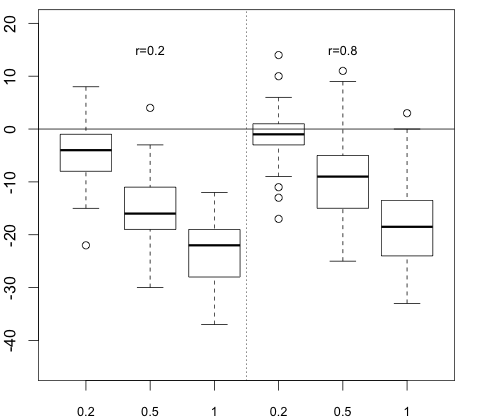}
\caption{$\widehat{K}-K^*$ obtained with $Q=\widehat{Q}$ (left) and $Q = Q^*$ (right) for
$\rho=(0.2,0.8)$ (denoted $r$ on the graph). {Box: first, second and third quartiles.}} \label{fig:SelectionK}
\end{center}
\end{figure}

\begin{figure}
\begin{center}
\includegraphics[width=0.45\textwidth]{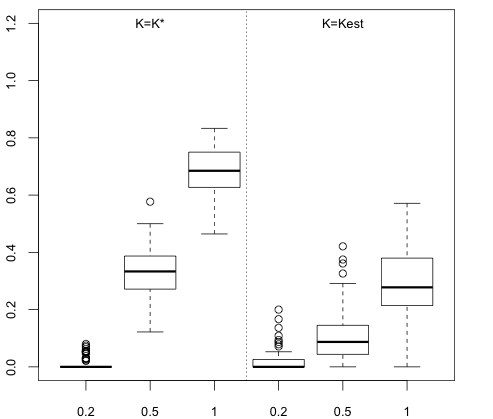}
\includegraphics[width=0.45\textwidth]{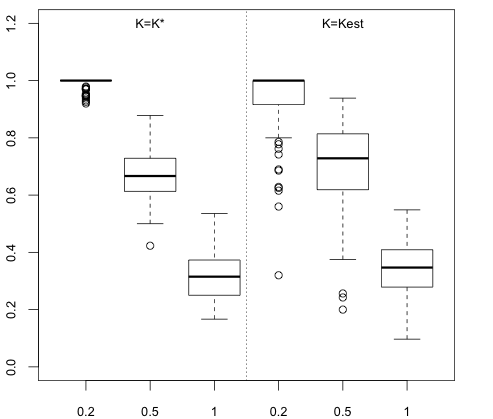}
\caption{FPR (left) and
TPR (right) with $Q=\widehat{Q}$, $\rho=0.8$ for $K=K^*$ and
$K=\widehat{K}$. $x$-axis: $\sigma$. {Box: first, second and third quartiles.}} \label{fig:SelectionKb}
\end{center}
\end{figure}

\paragraph{Discussion on the selection of $K$} 
Figure \ref{fig:SelectionK} shows that, whatever the level of the dependence, when the noise is small ($\sigma=0.2$), the selected number of segments is close to the true one and the breakpoints are well positioned (see Figure \ref{fig:SelectionKb}). When the detection problem gets difficult ($\sigma$ large), the selection procedure tends to underestimate the number of segments leading to a better precision on the breakpoints positioning compared to the true number of segments (smaller FPR). 

\paragraph{Discussion on the selection of $Q$}

The number of factors is strongly underestimated (see Table \ref{SelectionQ}) (close to $1$ in average for $\rho=0.2$ and for the different values of $\sigma$, not shown) meaning that only few factors are needed to capture the dependence structure. This underestimation does not alter much the estimation of $\Sigmabf$ nor the choice of the number of segments, represented in Figure \ref{fig:SelectionK}, when compared to the true number of factors. This underestimation acts as a regularization and that turns out to increase the power of procedure in terms of breakpoint positioning (in terms of FPR and TPR).

Moreover, the selected number of factors decreases slightly with the difficulty of the detection problem (with $\sigma$) leading to a decreasing precision of the estimation of $\Sigmabf$.

\begin{table}[!h]
\begin{center}
\begin{tabular}{|c|ccc|ccc|}\hline
\em & \multicolumn{3}{c|}{$(\widehat{Q},\widehat{K})$}&
\multicolumn{3}{c|}{$(Q^*,\widehat{K})$}\\ \hline $\sigma$ & 0.2 &
0.5 & 1 & 0.2
&  0.5 & 1 \\ \hline \hline mean of $Q$ & 3.37 & 2.74 & 2.39 &  &  9 &  \\
\hline $\RMSE(\Sigmabf)$ & 0.005  & 0.032 & 0.119  &
0.0048 &  0.032 & 0.124   \\
\hline $FPR$ & 0.016  & 0.110 &  0.288 &
0.039  &  0.175  &  0.413  \\
\hline $TPR$ & 0.93 &  0.69  & 0.34 &
0.93 &  0.597  & 0.262  \\
\hline
\end{tabular}
 \caption{\label{SelectionQ} Mean of $Q$,
$\RMSE(\Sigmabf)$, FPR and TPR for $\rho=0.8$ where $Q^*$ is the true
number of factors.}
\end{center}
\end{table}

\paragraph{Confounding between $\widehat{K}$ and $\widehat{\Sigmabf}$}
As observed in Figure \ref{fig:SelectionKb}, the true number of segments $K=K^*$ leads to numerous false positive breakpoints. In some sense, $K^*$ is too large for the data at hand. The consequence of this over-segmentation is that no factor are selected for $\sigma=0.5,1$, meaning that the dependence is captured by the segmentation (results not shown).

\begin{figure}
\begin{center}
\includegraphics[width=0.3\textwidth]{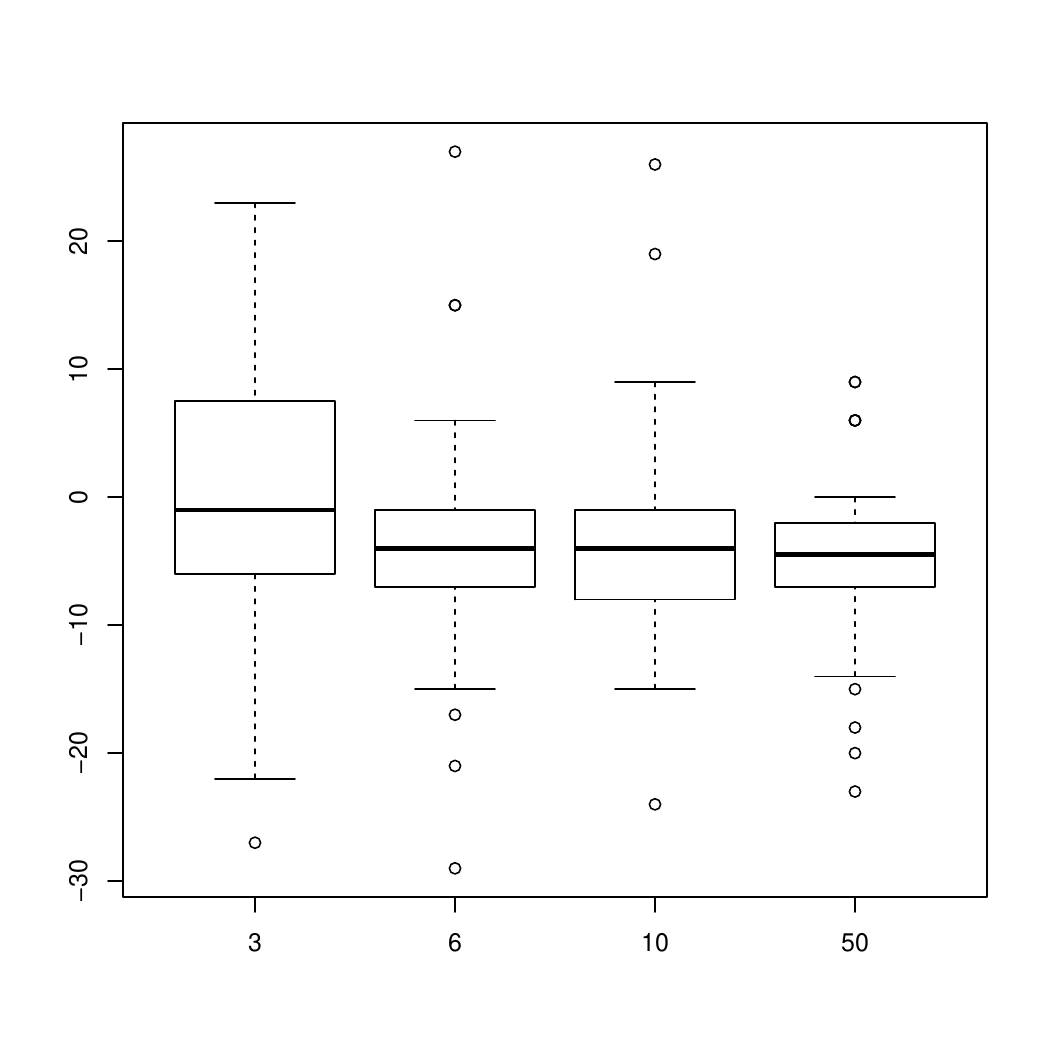}
\includegraphics[width=0.3\textwidth]{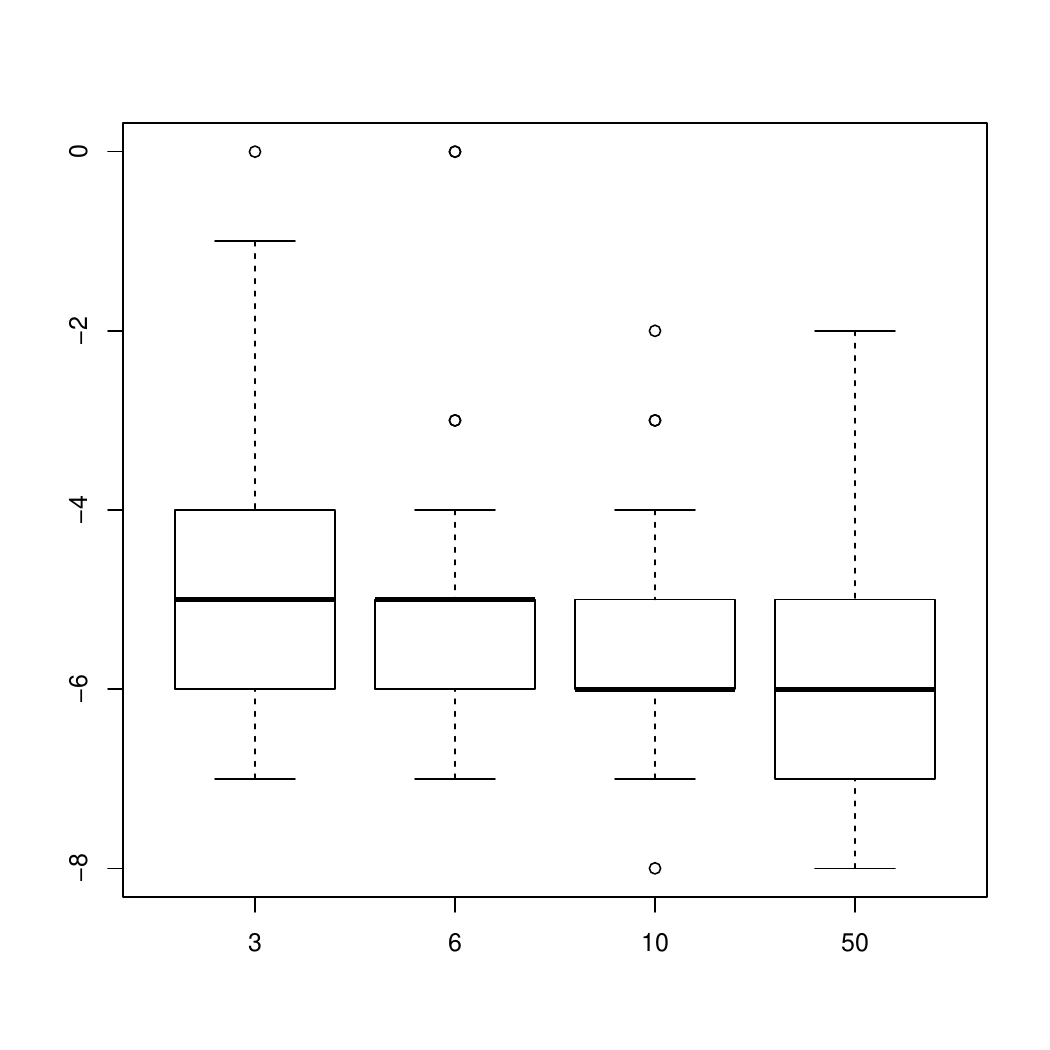}
\includegraphics[width=0.3\textwidth]{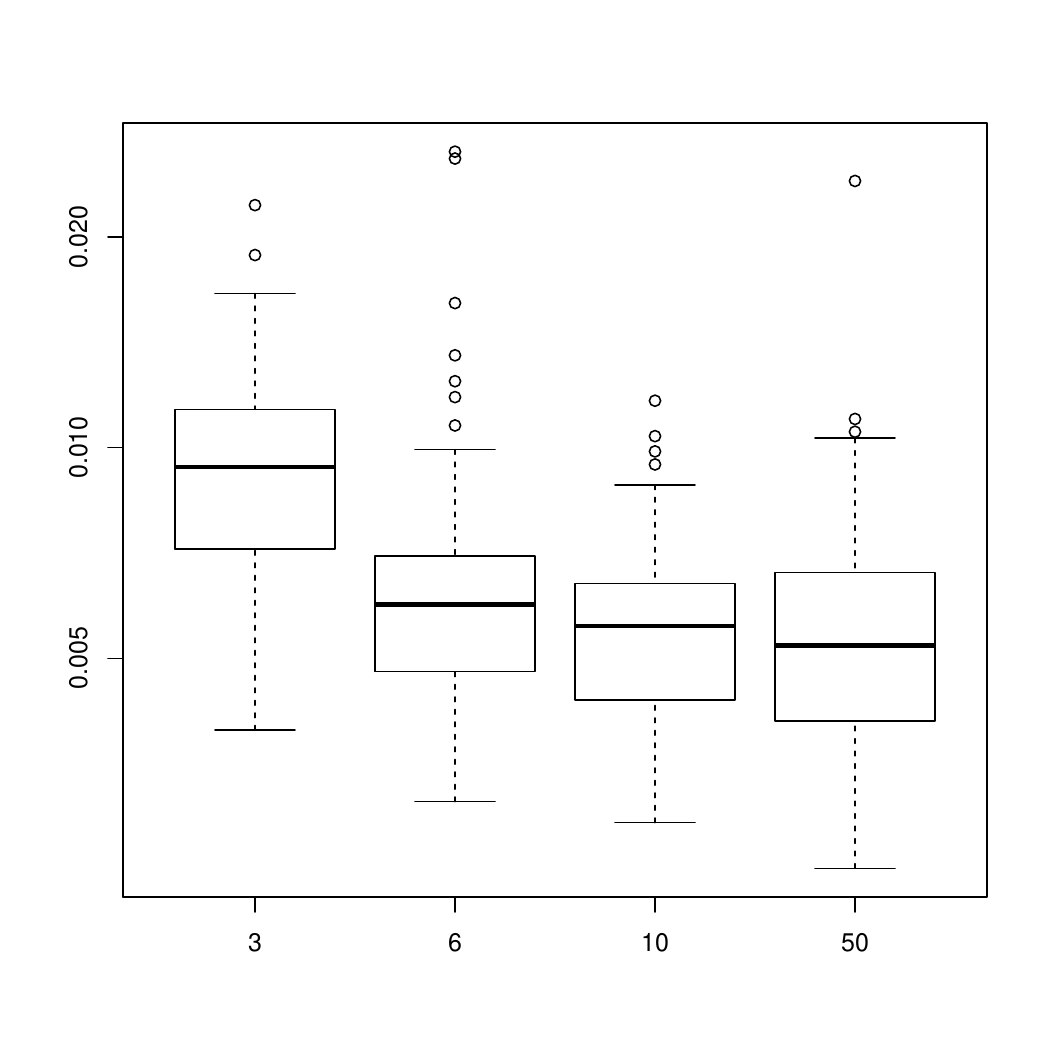} \\
\includegraphics[width=0.3\textwidth]{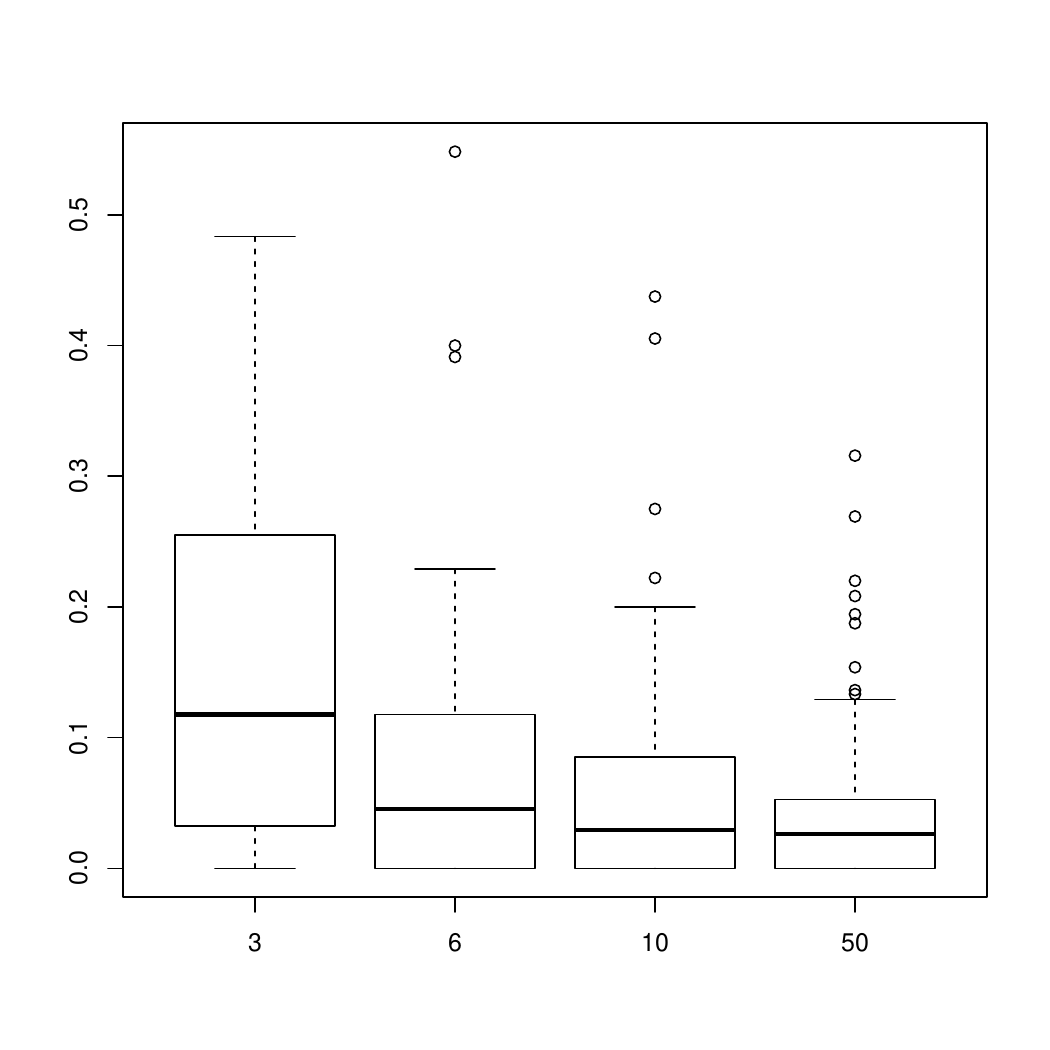}
\includegraphics[width=0.3\textwidth]{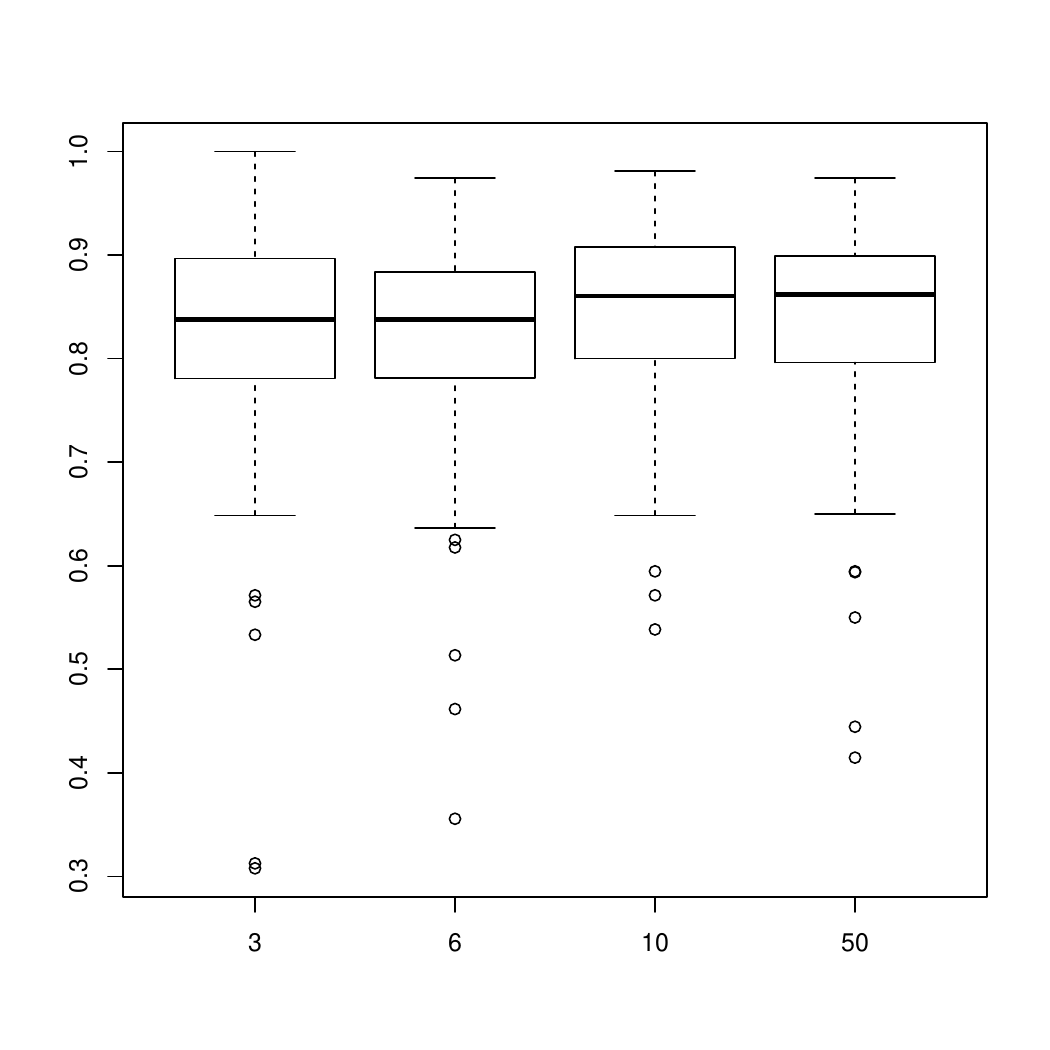}
\includegraphics[width=0.3\textwidth]{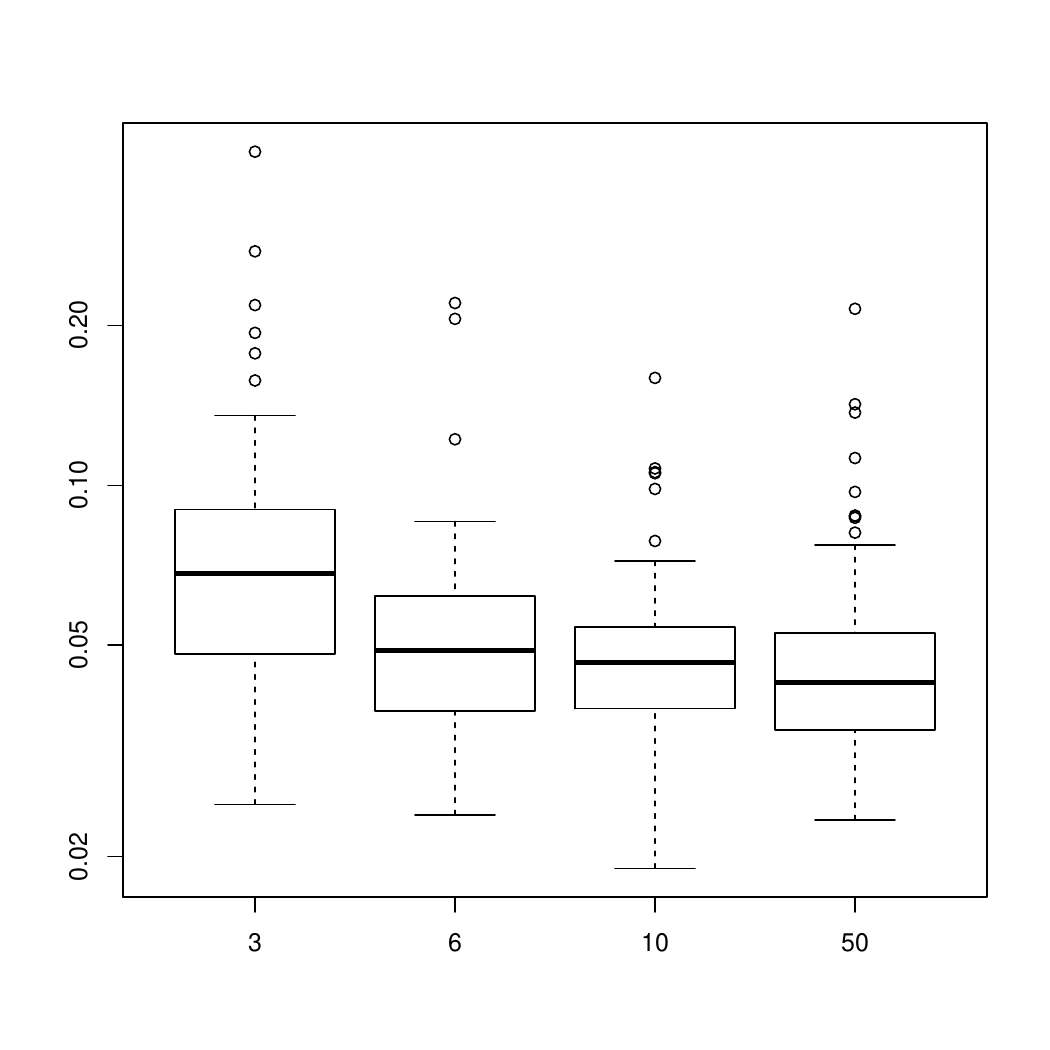}
\caption{Student simulation ($i$): Boxplots. Top: $\widehat{K} - K^*$ (left), $\widehat{Q} - Q^*$ (center),
RMSE($\Sigmabf$) (right). Bottom:  FPR (left), TPR (center), RMSE($\Tbf\mubf$) (right).
\label{fig:Student}}
\end{center}
\end{figure}

\begin{figure}
\begin{center}
\includegraphics[width=0.3\textwidth]{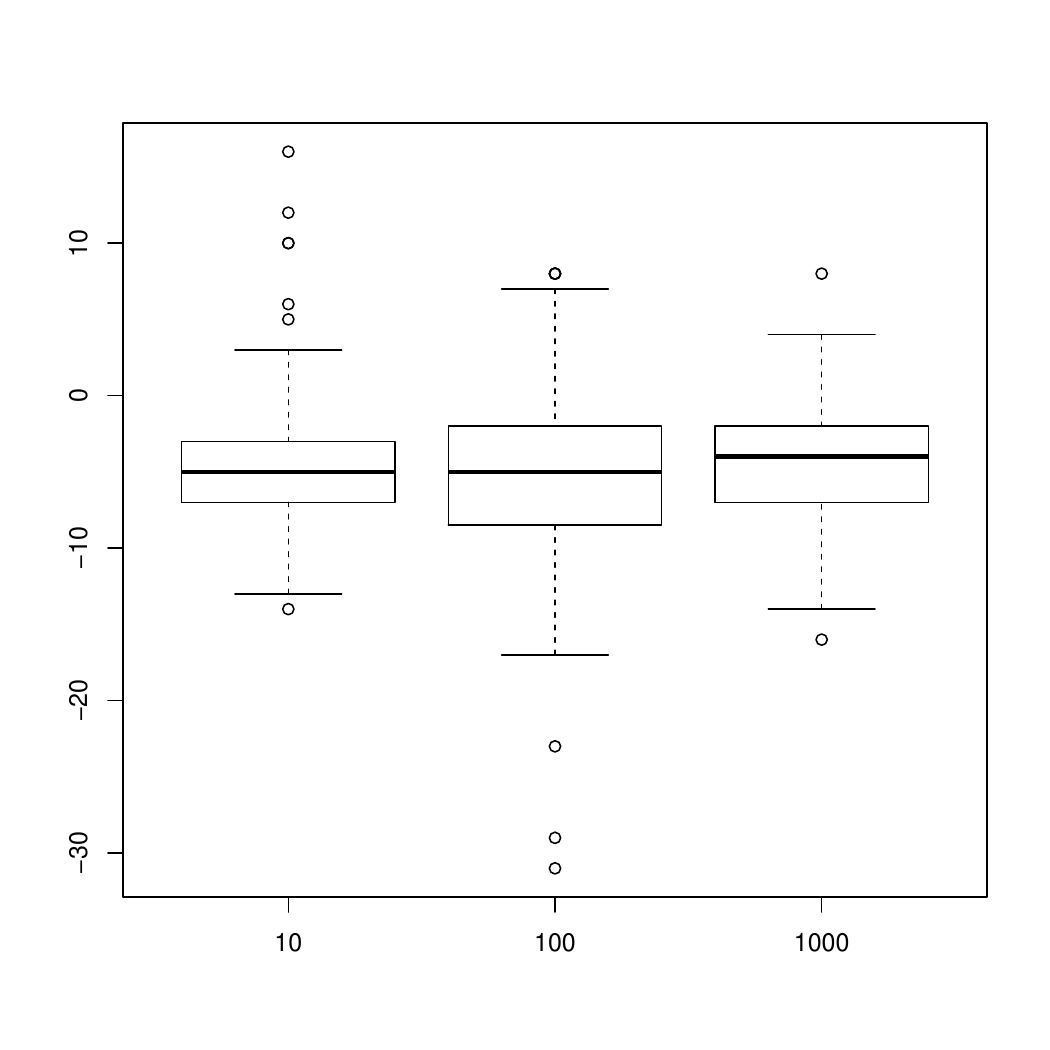}
\includegraphics[width=0.3\textwidth]{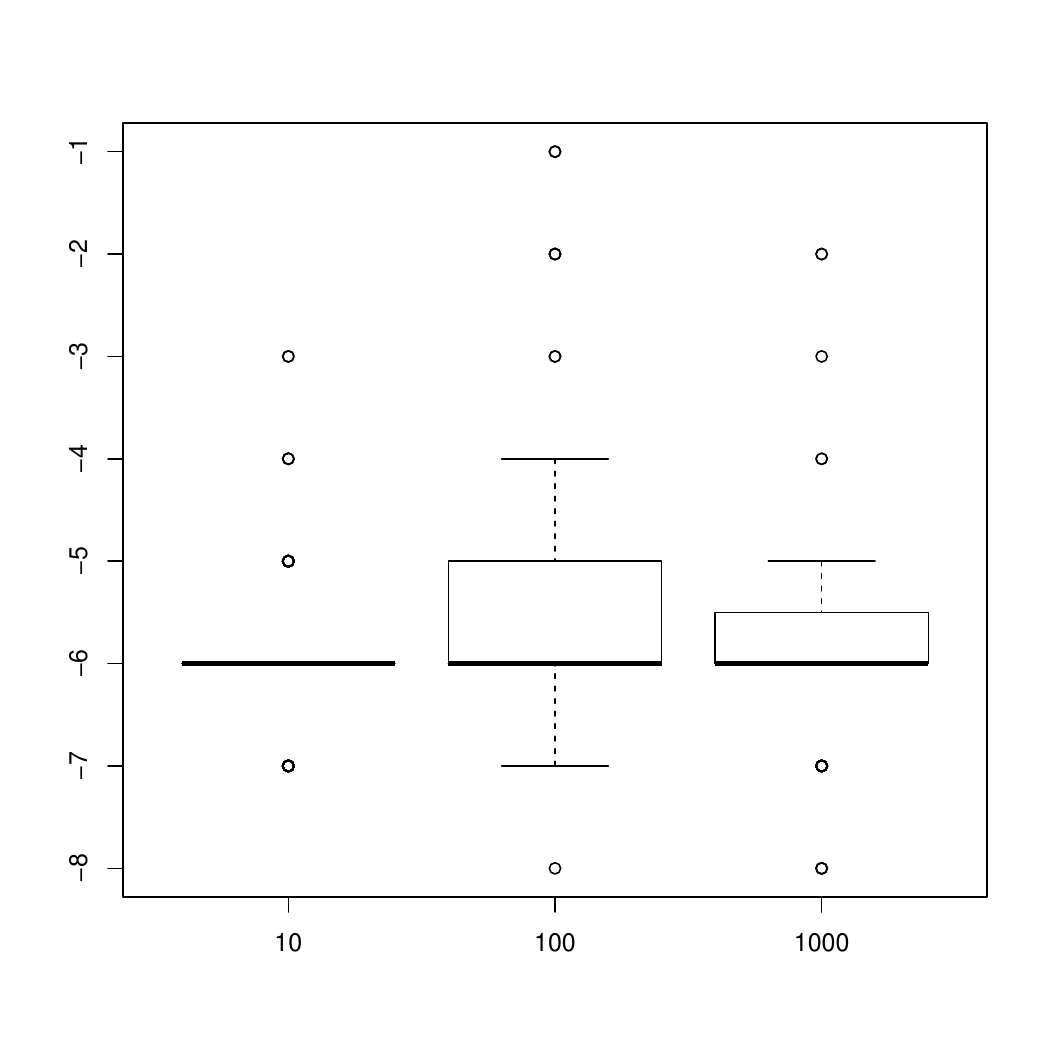}
\includegraphics[width=0.3\textwidth]{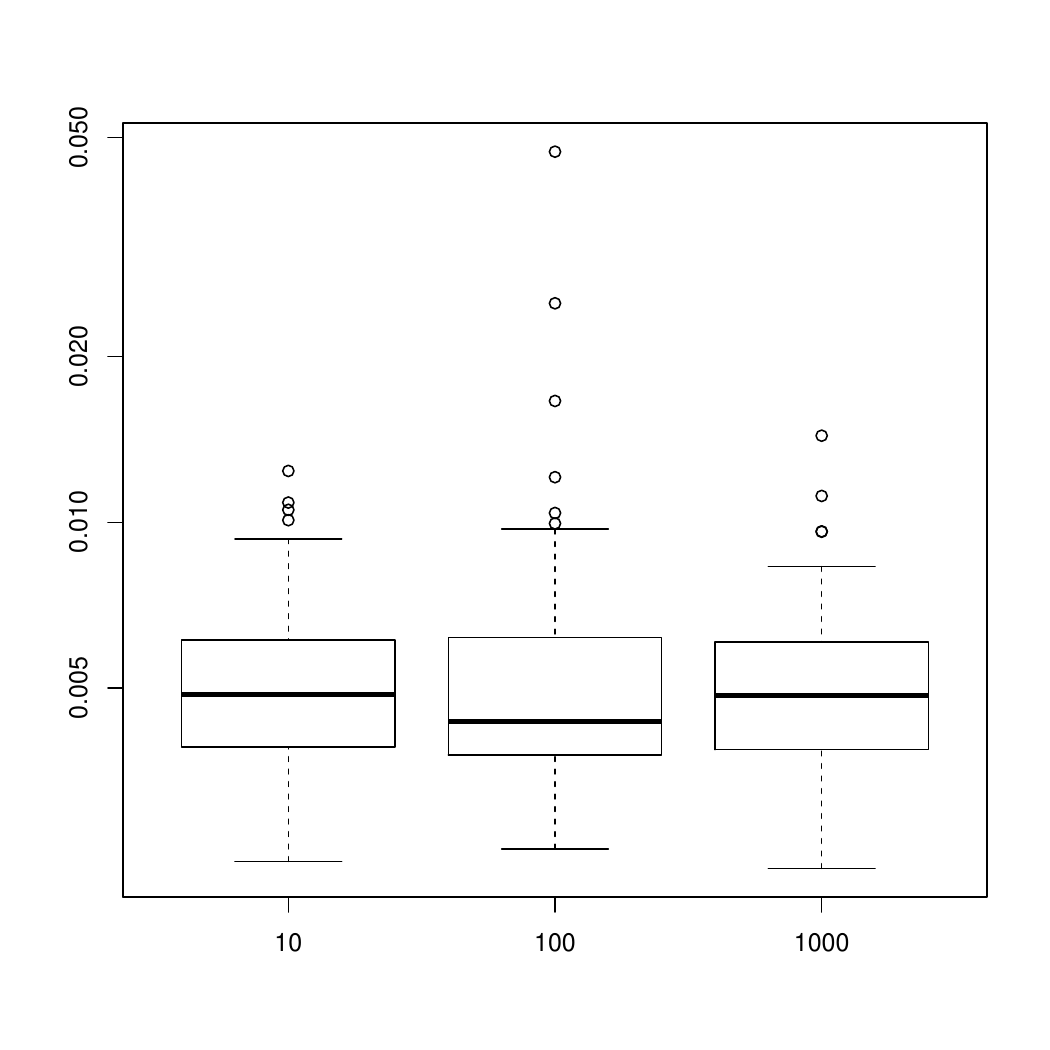} \\
\includegraphics[width=0.3\textwidth]{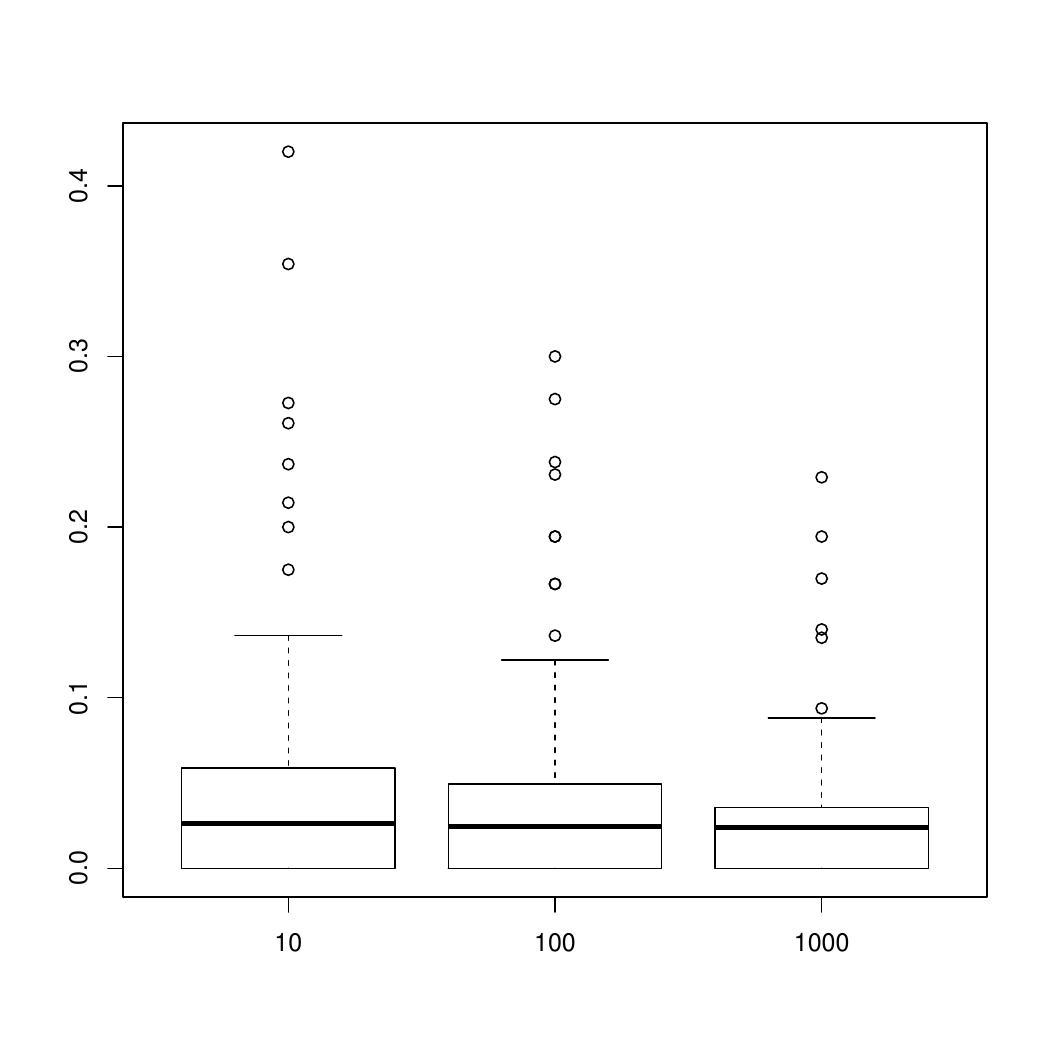}
\includegraphics[width=0.3\textwidth]{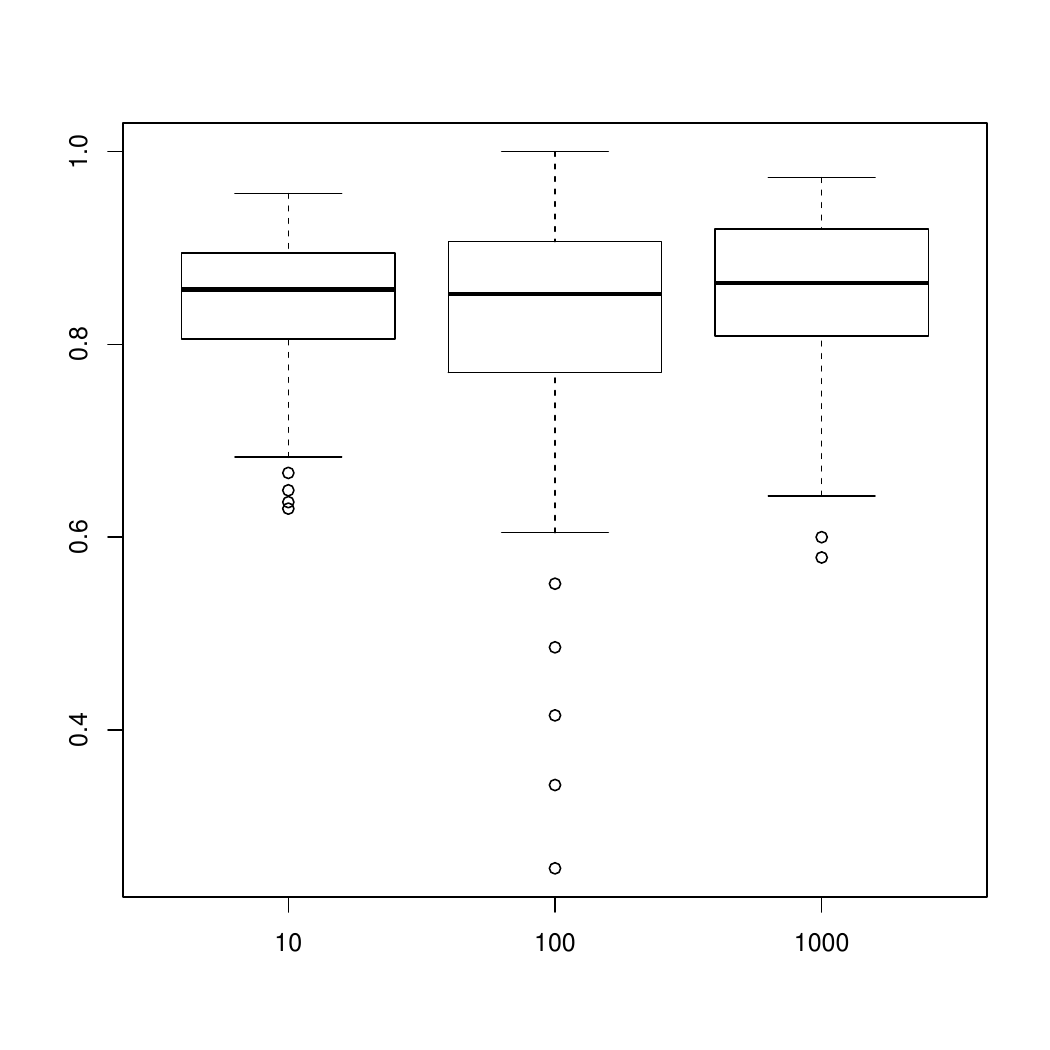}
\includegraphics[width=0.3\textwidth]{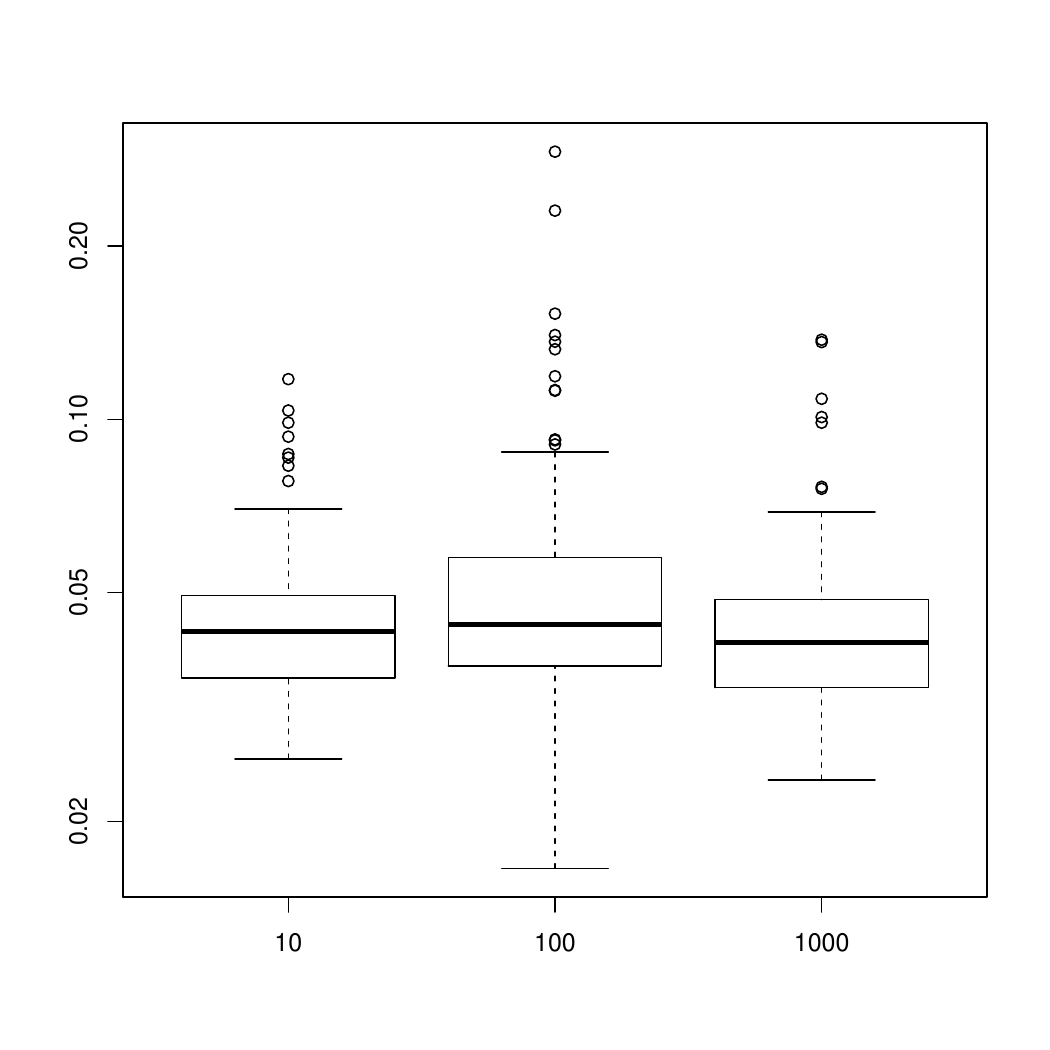}
\caption{Wishart simulation ($ii$): same legend as in Figure \ref{fig:Student}.
\label{fig:Wishart}}
\end{center}
\end{figure}

\paragraph{Robustness studies} 
Figure \ref{fig:Student} displays the results of the Student simulation ($i$), which is about the robustness to non-normality. We observe that the results are quite stable when the degree of freedom (df) varies in terms of $\widehat{K}$, $\widehat{Q}$ and TPR. In the extreme case df=3, as expected, the estimated number of segments is greater than in the other cases, resulting in a higher FPR. The other consequence is a slight degradation in the quality of the estimation of both $\Sigmabf$ and $\Tbf\mubf$. The results of the Wishart simulation ($ii$) about heterogeneous covariances are given in Figure \ref{fig:Wishart}. The results are even  more stable than in the Student simulation. \\
Overall, these simulations indicate a good behavior of the proposed procedure in presence of a reasonable deviation from normality and variance homogeneity.

\section{Application} \label{sec:appli}

\paragraph{Data description}
Scientific permanent Global Navigation Satellite Systems (GNSS) instruments continuously track electromagnetic signals from GPS satellites. Their data are generally collected by scientific, private or public services in near-real time or after getting a few days of observations to derive accurate coordinates. These coordinates are used to determine precise velocities of points located on the crust that constrain tectonic models and Earth's crust/mantle parameters, to infer mass transfer in the fluid layers of the Earth or for positioning applications (terrestrial reference frame).

We consider here the coordinate time series from five GNSS stations, located in the Michigan and Ohio States in USA. They were computed by the Jet Propulsion Laboratory (JPL) and can be downloaded at \url{sideshow.jpl.nasa.gov/post/series.html}. The stations are labeled MPLE, ADRI, BAYR, BRIG and DEFI. Only the longitude component has been investigated here: from the February 2002 to June 2013, 3776 longitude coordinates are available per station. Because they are separated by less than 250 km, their coordinate time series show similarities related to common ground deformation and correlated processing errors. The predominant effect in this component is the tectonic motion of the North American plate, which is about -16 mm/yr. When making difference of coordinate series from close stations, we can hope that this effect is canceled. Only residual differential motion as well as residual noise still remains in addition to sudden changes often related to equipment or environmental changes in one of the two stations. In this example, we use the station MPLE as the reference series and form four time series from the four other slave series, denoted ADRI-MPLE, BAYR-MPLE, BRIG-MPLE and DEFI-MPLE. \\
In addition to the coordinate series, we also have access to some of recorded changes that occurred in each station (see Table \ref{tab:known-changes}). These known events can be used to partially validate the inferred breakpoints.

\begin{figure}
\begin{center}
\includegraphics[width=1\textwidth]{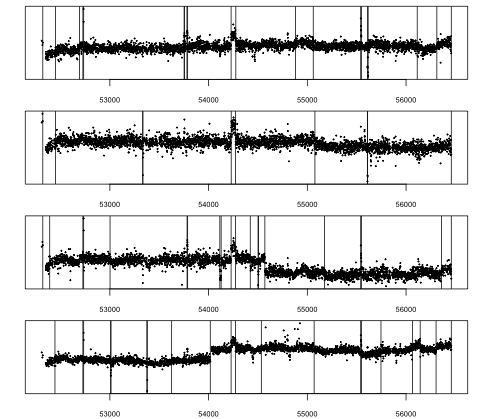}
\caption{Estimated breakpoints for the 4 series obtained with the
segmentation only. {From bottom to top, the series are ADRI-MPLE, BAYR-MPLE, BRIG-MPLE and DEFI-MPLE}.} \label{fig:ResSegSeule}
\end{center}
\end{figure}
\begin{figure}
\begin{center}
\includegraphics[width=1\textwidth]{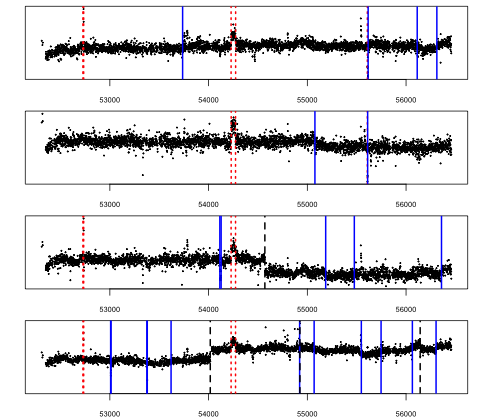}
\caption{Estimated breakpoints for the 4 series when the
dependence is taken into account. Dotted lines (in red): common
breakpoints among almost the 4 series. Solid line (in blue):
series-specific breakpoints. Dashed lines (in black): known
series-specific breakpoints. {From bottom to top, the series are ADRI-MPLE, BAYR-MPLE, BRIG-MPLE and DEFI-MPLE}.} \label{fig:ResSeg}
\end{center}
\end{figure}

\paragraph{Accounting for dependence}
The proposed model selection procedure leads to select $\widehat{Q}=1$ factor, which means that a dependence between the series does exist. The same procedure selects $\widehat{K}=46$ segments. When the series are assumed to be independent ($Q=0$), the selected number of segments $70$, which is significantly larger. Figures \ref{fig:ResSeg} and \ref{fig:ResSegSeule} give the four series with the estimated breakpoint positions (vertical lines) obtained with and without dependence. The comparison with Table \ref{tab:known-changes} shows that the breakpoints arising when the dependence is omitted do not correspond to known events. This suggest that the inclusion of dependence avoids false positive detections.

\paragraph{Breakpoint interpretation}
Two categories can be distinguished among the estimated breakpoints.
\begin{itemize}
\item The breakpoints being common to most of the four series (Figure \ref{fig:ResSeg}, red dotted lines): these breakpoints can be due to changes in the reference series MPLE. The one at day 54228 is known to be due to antenna and receiver changes (see Table \ref{tab:known-changes}).
\item The series-specific breakpoints: some of them are due to known instrumental changes (black dashed lines). The change at day 53620 in series 4 (DEFI-MPLE) can be due to the known change in MPLE (see Table \ref{tab:known-changes}). The remaining ones (black solid lines) are still unexplained, even if some are close to known events.
\end{itemize}
Although it is not referred as known event, the change detected in series 2 (BAYR-MPLE) at day 55077 but is very well marked. Series 4 (DEFI-MPLE) displays a periodic behavior, which is compensated by a large number of estimated breakpoints. This behavior is likely to result from the large distance between stations BAYR and MPLE: because of this large distance, the tectonic motion part is not completely corrected by the difference.

\begin{table}
\begin{center}
{\small
\begin{tabular}{lp{.3\textwidth}}
Serie & date (time in week) \\
\hline
ADRI &
2004-12-02~(53341)
2005-09-15~(53628)
2006-08-02~(53949) \\
\hline
BAYR & 
2004-12-02~(53341) 
2005-08-31~(53613) 
2006-07-11~(53927) \\
\hline
BRIG & 
2004-12-02~(53341) 
2005-09-12~(53625) 
2008-04-14~(54570$^a$) \\
\hline
DEFI &
2006-10-11~(54019$^a$)
2009-04-07~(54928)
2011-01-26~(55587)
2012-08-02~(56141$^a$) 
2012-11-05~(56236$^a$) \\
\hline
MPLE &
2004-12-02~(53341)
2005-09-14~(53627)
2006-12-28~(54097)
2007-05-08~(54228$^b$)
2007-10-05~(54378) \\
\hline
\end{tabular}
}
\end{center}
\caption{Known changes in the five series. All changes corresponds to a change of receiver, except ($^a$): change of antenna and ($^b$) change of both receiver and antenna.}
\label{tab:known-changes}
\end{table}

\begin{figure}
\begin{center}
\includegraphics[width=1\textwidth]{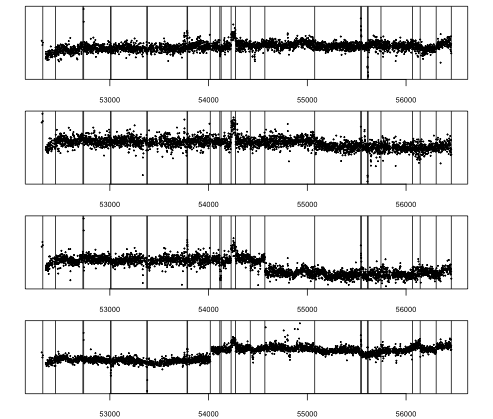}
\caption{Estimated breakpoints for the 4 series obtained with the simultaneous segmentation. {From bottom to top, the series are ADRI-MPLE, BAYR-MPLE, BRIG-MPLE and DEFI-MPLE}.} \label{fig:ResSegSimultanee}
\end{center}
\end{figure}

\paragraph{Simultaneous segmentation}
Because of the presence of common breakpoints, we applied a simultaneous segmentation (with between-series correlation) to the four series. The results are given in Figure \ref{fig:ResSegSimultanee}. The proposed procedure selects $\widehat{Q}=2$ hidden factors and $\widehat{K}=27$ common segments. A visual inspection (and a comparison with Figures \ref{fig:ResSegSeule} and \ref{fig:ResSeg}) reveals that some are shared by at least two series and several others are only obvious in one series (especially the last one). Simultaneous segmentation does not provide any insight to distinguish between common and series-specific breakpoints.

\begin{figure}
\begin{center}
\includegraphics[width=0.4\textwidth, clip=]{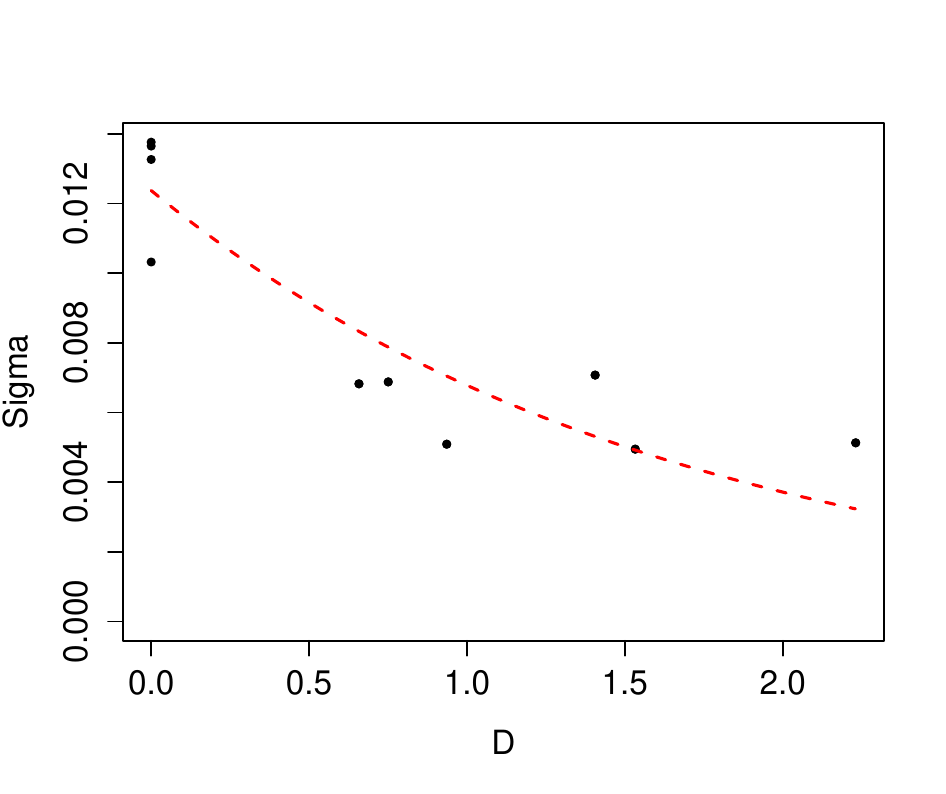}
\qquad
\includegraphics[width=0.3\textwidth, clip=]{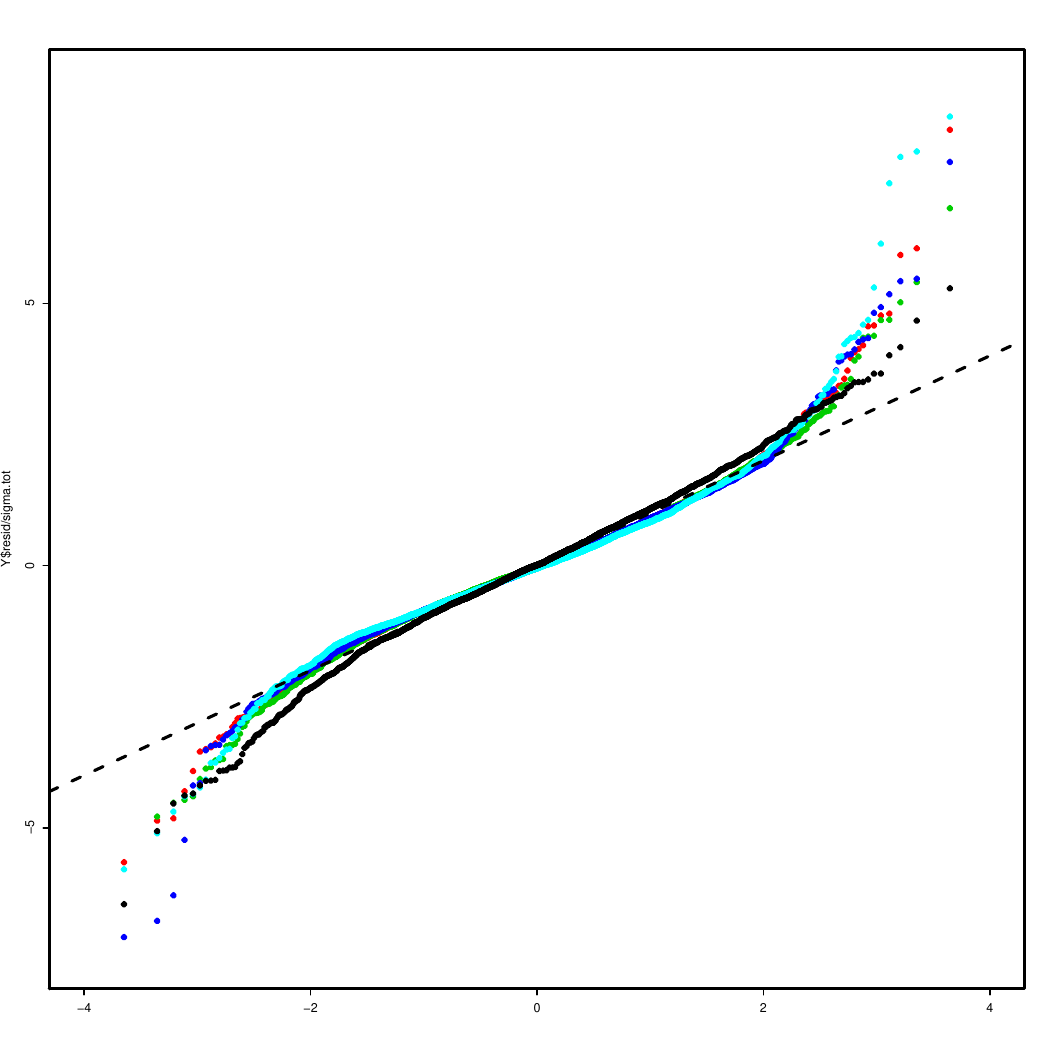}
\caption{Left: Estimation of $\Sigmabf$ according to the distance between
series. Right: Normal QQ-plot of the residual (one color per series, black=random Student sample with df=10).} \label{fig:SigmaDistance}
\end{center}
\end{figure}

\paragraph{Covariance structure}
The choice $\widehat{Q}=1$ seems to be sufficient to capture the dependence structure among series. $\Sigmabf$ well captures the spatial dependence between series as shown in 
Figure \ref{fig:SigmaDistance} (left). We observe that the estimated correlation decreases as the distance increases, although this structure has not been imposed when estimating $\Sigmabf$.

\paragraph{Goodness of fit} 
To assess the goodness of fit of our model, we checked the distribution of the residuals (see Figure \ref{fig:SigmaDistance} (right)). We observe a departure from the normality, similar to this of a Student distribution with about 10 degrees of freedom. The Student simulation study ($i$) from Section \ref{sec:simu} suggests that the proposed methodology is robust to this type of departure.

\section{Discussion}

\paragraph{Joint segmentation procedure}
We proposed a comprehensive framework for maximum likelihood segmentation of multiple series in presence of between-series correlation. The procedure involves an EM algorithm for parameter inference and a model selection procedure for choosing the number of hidden factors and the number of segments. The use of a factor model allows the use of the Dynamic Programming algorithm, which results in an efficient algorithm. The time-efficiency could probably be improved, combining the two-stage DP algorithm from \cite{PLB11,PLHRTR11} with the pruned version of DP proposed by \cite{G15}. The combination with improved version from \cite{killick_pelt} would be more complex as this method embeds the selection of the number of segments.

\paragraph{Covariance regularization}
The factor model is not only useful to design an efficient algorithm. It also acts as a regularization for the estimation of the between-series covariance matrix. We showed that this regularization significantly improves the breakpoint detection. In the simulation study, the selected number of factors is drastically reduced with respect to the true one. This regularization goes along with a similar number of true breakpoint detections and a reduced number of false detections.

\paragraph{Within-series dependence}
The factor model we consider enables us to break the between-series dependence but not the within-series dependence. The proposed model does not account for such a dependence, although it may exist. Still, in the one-series segmentation case, the least-square estimate (that is equivalent to maximum-likelihood for an independent Gaussian signal) has been shown to be consistent, even in presence of long-range dependence (see \cite{LaM00}). \SR{}{However, in the non-asymptotic framework, time dependence may affect the segmentation results. Recently, \cite{CLL17} proposed a two-stage whitening strategy that allows the use of Dynamic Programming preserving the statistical guaranties.} The introduction of time dependence in our model will be considered in a further work. 

\paragraph{Acknowledgements} This work has been partly supported by mathamsud 16-MATH-03 SIDRE project. 

\bibliography{modlin}

\begin{thebibliography}{Tototo}

\bibitem[Amiri-Simkooei (2009)]{Amiri2009}%
{\sc Amiri-Simkooei, A.~R.}
\newblock (2009).
\newblock Noise in multivariate gps position time-series.
\newblock {\em Journal of Geodesy}.
\newblock {\bf 83} 175--187.

\bibitem[Arlot {\em et~al.} (2016)]{ACH2016}%
{\sc Arlot, S.}, {\sc Celisse, A.} and {\sc Harchaoui, Z.}
\newblock (2016).
\newblock A kernel multiple change-point algorithm via model selection.

\bibitem[Aue {\em et~al.} (2009)]{AHHR2009}%
{\sc Aue, A.}, {\sc H\"ormann, S.}, {\sc Horv\'ath, L.},  and {\sc Reimherr,
  M.}
\newblock (2009).
\newblock Break detection in the covariance structure of multivariate time
  series models.
\newblock {\em Annals of Statistics}.
\newblock {\bf 37} 4046–4087.

\bibitem[Bai and Perron (2003)]{BP03}%
{\sc Bai, J.} and {\sc Perron, P.}
\newblock (2003).
\newblock Computation and analysis of multiple structural change models.
\newblock {\em J. Appl. Econ.}
\newblock {\bf 18} 1--22.

\bibitem[Bai and Ng (2002)]{BaN02}%
{\sc Bai, J.} and {\sc Ng, S.}
\newblock (2002).
\newblock Determining the number of factors in approximate factor models.
\newblock {\em Econometrica}.
\newblock {\bf 70}~{\bf (1)} 191--221.

\bibitem[{Barigozzi} {\em et~al.} (2016)]{BCF17}%
{\sc {Barigozzi}, M.}, {\sc {Cho}, H.} and {\sc {Fryzlewicz}, P.}
\newblock (2016), {Simultaneous multiple change-point and factor analysis for
  high-dimensional time series}.
\newblock Technical report, arXiv:1612.06928.

\bibitem[Bellman (1961)]{bellman_approximation_1961}%
{\sc Bellman, R.}
\newblock (1961).
\newblock On the approximation of curves by line segments using dynamic
  programming.
\newblock {\em Commun. {ACM}}.
\newblock {\bf 4}~{\bf (6)} 284.

\bibitem[Cabrippeto {\em et~al.} (2016)]{Cabri2016}%
{\sc Cabrippeto, J.}, {\sc Tuerlinckx, F.}, {\sc Kuppens, P.}, {\sc Grassmann,
  M.} and {\sc Ceulemans, E.}
\newblock (2016).
\newblock Detecting correlation changes in multivariate time series: A
  comparison of four non-parametric change-point detection methods.
\newblock In {\em Behavior Research Methods}, (M.~Jones,~ed.), 1--18. Springer.

\bibitem[Caussinus and Mestre (2004)]{CM04}%
{\sc Caussinus, H.} and {\sc Mestre, O.}
\newblock (2004).
\newblock Detection and correction of artificial shifts in climate series.
\newblock {\em JRSS-C}.
\newblock {\bf 53}~{\bf (3)} 405--425.

\bibitem[Chakar {\em et~al.} (2017)]{CLL17}%
{\sc Chakar, S.}, {\sc Lebarbier, E.}, {\sc L{\'e}vy-Leduc, C.}, {\sc Robin,
  S.} {\em et~al.}
\newblock (2017).
\newblock A robust approach for estimating change-points in the mean of an
  {A}{R}(1) process.
\newblock {\em Bernoulli}.
\newblock {\bf 23}~{\bf (2)} 1408--1447.

\bibitem[Cho and Fryzlewicz (2012)]{CF2012}%
{\sc Cho, H.} and {\sc Fryzlewicz, P.}
\newblock (2012).
\newblock Multiscale and multilevel technique for consistent segmentation of
  nonstationary time series.
\newblock {\em Statistica Sinica}.
\newblock {\bf 22} 207--229.

\bibitem[Cho and Fryzlewicz (2015)]{ChF15}%
{\sc Cho, H.} and {\sc Fryzlewicz, P.}
\newblock (2015).
\newblock Multiple-change-point detection for high dimensional time series via
  sparsified binary segmentation.
\newblock {\em Journal of the Royal Statistical Society: Series B (Statistical
  Methodology)}.
\newblock {\bf 77}~{\bf (2)} 475--507.

\bibitem[Dempster {\em et~al.} (1977)]{DLR77}%
{\sc Dempster, A.~P.}, {\sc Laird, N.~M.} and {\sc Rubin, D.~B.}
\newblock (1977).
\newblock Maximum likelihood from incomplete data via the {EM} algorithm.
\newblock {\em Journal of the Royal Statistical Society Series B}.
\newblock {\bf 39} 1--38.

\bibitem[Dobigeon and Tourneret (2007)]{DT2007}%
{\sc Dobigeon, N.} and {\sc Tourneret, J.}
\newblock (2007).
\newblock Joint segmentation of piecewise constant autoregressive processes by
  using a hierarchical model and a bayesian sampling approach.
\newblock {\em IEEE Transactions on Signal Processing}.
\newblock {\bf 55}~{\bf (4)} 1251--1263.

\bibitem[Dong {\em et~al.} (2006)]{D2006}%
{\sc Dong, D.}, {\sc Fang, P.}, {\sc Bock, Y.}, {\sc Webb, F.}, {\sc
  Prawirodirdjo, L.}, {\sc Kedar, S.},  and {\sc Jamason, P.}
\newblock (2006).
\newblock Spatiotemporal filtering using principal component analysis and
  karhunen-loeve expansion approaches for regional gps network analysis.
\newblock {\em Journal of Geophysical Research (Solid Earth)}.
\newblock  111--3405.

\bibitem[van Dyk (2000)]{VD00}%
{\sc van Dyk, D.}
\newblock (2000).
\newblock Fitting mixed-effects models using efficient {EM}-type algorithms.
\newblock {\em Jour. Comp. and Graph. Statistics}.
\newblock {\bf 9} 78--98.

\bibitem[Friguet {\em et~al.} (2009)]{FKC09}%
{\sc Friguet, C.}, {\sc Kloareg, M.} and {\sc Causeur, D.}
\newblock (2009).
\newblock A factor model approach to multiple testing under dependence.
\newblock {\em J. Amer. Statist. Assoc.}
\newblock {\bf 488} 1406--15.

\bibitem[Gazeaux {\em et~al.} (2015)]{gazeaux2015joint}%
{\sc Gazeaux, J.}, {\sc Lebarbier, E.}, {\sc Collilieux, X.} and {\sc
  M{\'e}tivier, L.}
\newblock (2015).
\newblock Joint segmentation of multiple {GPS} coordinate series.
\newblock {\em Journal de la Soci{\'e}t{\'e} Fran{\c{c}}aise de Statistique}.
\newblock {\bf 156}~{\bf (4)} 163--179.

\bibitem[Harle {\em et~al.} (2016)]{HCGA2016}%
{\sc Harle, F.}, {\sc Chatelain, F.}, {\sc Gouy-Pailler, C.} and {\sc Achard,
  S.}
\newblock (2016).
\newblock Bayesian model for multiple change-points detection in multivariate
  time series.
\newblock {\em IEEE Transactions on Signal Processing}.
\newblock {\bf 64}~{\bf (16)} 4351--4362.

\bibitem[Killick {\em et~al.} (2012)]{killick_pelt}%
{\sc Killick, R.}, {\sc Fearnhead, P.} and {\sc Eckley, I.}
\newblock (2012).
\newblock Optimal detection of changepoints with a linear computational cost.
\newblock {\em Journal of the American Statistical Association}.
\newblock {\bf 107}~{\bf (500)} 1590--1598.

\bibitem[Lai {\em et~al.} (2005)]{LJK05}%
{\sc Lai, W.}, {\sc Johnson, M.}, {\sc Kucherlapati, R.} and {\sc Park, P.~J.}
\newblock (2005).
\newblock Comparative analysis of algorithms for identifying amplifications and
  deletions in array {CGH} data.
\newblock {\em Bioinformatics}.
\newblock {\bf 0}~{\bf (0)} 1--8.

\bibitem[Lauritzen (1996)]{Lau96}%
{\sc Lauritzen, S.~L.}
\newblock (1996).
\newblock {\em Graphical Models}.
\newblock Oxford Statistical Science Series. Clarendon Press.

\bibitem[Lavielle (2005)]{Lav05}%
{\sc Lavielle, M.}
\newblock (2005).
\newblock Using penalized contrasts for the change-point problem.
\newblock {\em Signal Processing}.
\newblock {\bf 85}~{\bf (8)} 1501--1510.

\bibitem[Lavielle and Moulines (2000)]{LaM00}%
{\sc Lavielle, M.} and {\sc Moulines, E.}
\newblock (2000).
\newblock Least-squares estimation of an unknown number of shifts in a time
  series.
\newblock {\em Journal of time series analysis}.
\newblock {\bf 21}~{\bf (1)} 33--59.

\bibitem[Lavielle and Teyssiere (2006)]{LT2006}%
{\sc Lavielle, M.} and {\sc Teyssiere, G.}
\newblock (2006).
\newblock Detection of multiple change-points in multivariate time series.
\newblock {\em Lithuanian Mathematical Journal}.
\newblock {\bf 46} 287--306.

\bibitem[Lebarbier (2005)]{Leb05}%
{\sc Lebarbier, E.}
\newblock (2005).
\newblock Detecting multiple change-points in the mean of gaussian process by
  model selection.
\newblock {\em Signal Proc.}
\newblock {\bf 85} 717--36.

\bibitem[Lopes and West (2004)]{LoW04}%
{\sc Lopes, H.~F.} and {\sc West, M.}
\newblock (2004).
\newblock Bayesian model assessment in factor analysis.
\newblock {\em Statistica Sinica}.
\newblock {\bf 14}~{\bf (1)} 41--68.

\bibitem[Maidstone {\em et~al.} (2016)]{Maidstone2016}%
{\sc Maidstone, R.}, {\sc Hocking, T.}, {\sc Rigaill, G.} and {\sc Fearnhead,
  P.}
\newblock (2016).
\newblock On optimal multiple changepoint algorithms for large data.
\newblock {\em Statistics and Computing}.
\newblock  1--15.

\bibitem[Matteson and James (2014)]{MJ2014}%
{\sc Matteson, D.} and {\sc James, N.}
\newblock (2014).
\newblock A nonparametric approach for multiple change point analysis of
  multivariate data.
\newblock {\em Journal of the American Statistical Association}.
\newblock {\bf 109}~{\bf (505)} 334--345.

\bibitem[Mestre {\em et~al.} (2013)]{Mal13}%
{\sc Mestre, O.}, {\sc Domonkos, P.}, {\sc Picard, F.}, {\sc Auer, I.}, {\sc
  Robin, S.}, {\sc Lebarbier, E.}, {\sc Böhm, R.}, {\sc Aguilar, E.}, {\sc
  Guijarro, J.}, {\sc Vertachnik, G.}, {\sc Klancar, M.}, {\sc Dubuisson, B.}
  and {\sc Stepanek, P.}
\newblock (2013).
\newblock Homer : a homogenization software - methods and applications.
\newblock {\em Quarterly Journal of the Hungarian Meteorological Service}.
\newblock {\bf 117}~{\bf (1)} 47--67.

\bibitem[Nowak {\em et~al.} (2011)]{NHPT2011}%
{\sc Nowak, G.}, {\sc Hastie, T.}, {\sc Pollack, J.} and {\sc Tibshirani, R.}
\newblock (2011).
\newblock A fused lasso latent feature model for analyzing multi-sample acgh
  data.
\newblock {\em Biostatistics}.
\newblock {\bf 0}~{\bf (0)} 1–26.

\bibitem[Picard {\em et~al.} (2005)]{PRL05}%
{\sc Picard, F.}, {\sc Robin, S.}, {\sc Lavielle, M.}, {\sc Vaisse, C.} and
  {\sc Daudin, J.-J.}
\newblock (2005).
\newblock A statistical approach for array {CGH} data analysis.
\newblock {\em BMC Bioinformatics}.
\newblock {\bf 6}~{\bf (27)}~1.

\bibitem[Picard {\em et~al.} (2011a)]{PLHRTR11}%
{\sc Picard, F.}, {\sc Lebarbier, E.}, {\sc Hoebeke, M.}, {\sc Rigaill, G.},
  {\sc Thiam, B.} and {\sc Robin, S.}
\newblock (2011a).
\newblock Joint segmentation, calling and normalization of multiple cgh
  profiles.
\newblock {\em Biostatistics}.
\newblock {\bf 12}~{\bf (3)} 413--428.

\bibitem[Picard {\em et~al.} (2011b)]{PLB11}%
{\sc Picard, F.}, {\sc Lebarbier, E.}, {\sc Budinska, E.} and {\sc Robin, S.}
\newblock (2011b).
\newblock Joint segmentation of multivariate gaussian processes using mixed
  linear models.
\newblock {\em Comput. Statist. and Data Analysis}.
\newblock {\bf (2)} 1160--70.

\bibitem[Rigaill (2015)]{G15}%
{\sc Rigaill, G.}
\newblock (2015).
\newblock A pruned dynamic programming algorithm to recover the best
  segmentations in 1 to k{max} changes.
\newblock {\em numero special du Journal de la SFdS sur la detection de
  ruptures}.
\newblock {\bf 156}~{\bf (4)} 180--205.

\bibitem[Rubin and Thayer (1982)]{RuT82}%
{\sc Rubin, D.~B.} and {\sc Thayer, D.~T.}
\newblock (1982).
\newblock {EM} algorithms for ml factor analysis.
\newblock {\em Psychometrika}.
\newblock {\bf 47}~{\bf (1)} 69--76.

\bibitem[Vert and Bleakley (2010)]{VB2010}%
{\sc Vert, J.} and {\sc Bleakley, K.}
\newblock (2010).
\newblock Fast detection of multiple change-points shared by many signals using
  group lars.
\newblock {\em Advances in Neural Information Processing Systems}.
\newblock {\bf 23} 2343--2351.

\bibitem[Wang and Samworth (2018)]{WaS18}%
{\sc Wang, T.} and {\sc Samworth, R.~J.}
\newblock (2018).
\newblock High dimensional change point estimation via sparse projection.
\newblock {\em Journal of the Royal Statistical Society: Series B (Statistical
  Methodology)}.
\newblock {\bf 80}~{\bf (1)} 57--83.

\bibitem[Williams (2003)]{William2003}%
{\sc Williams, S.}
\newblock (2003).
\newblock Offsets in global positioning system time series.
\newblock {\em Journal of Geophysical Research: Solid Earth}.
\newblock {\bf 108} 2310.

\bibitem[Zhang and Siegmund (2007)]{ZhS07}%
{\sc Zhang, N.~R.} and {\sc Siegmund, D.~O.}
\newblock (2007).
\newblock A modified {B}ayes information criterion with applications to the
  analysis of comparative genomic hybridization data.
\newblock {\em Biometrics}.
\newblock {\bf 63}~{\bf (1)} 22--32.

\bibitem[Zhang and Siegmund (2008)]{ZS2008}%
{\sc Zhang, N.} and {\sc Siegmund, O.}
\newblock (2008).
\newblock Detecting simultaneous change-points in multiple sequences.
\newblock {\em Biometrika}.
\newblock {\bf 0}~{\bf (0)} 1–18.

\end{thebibliography}
\bibliographystyle{astats}

\end{document}